\documentclass[aps,prb,twocolumn,floatfix,showpacs,prb]{revtex4-1} 
\usepackage{graphicx}
\usepackage{amsmath}
\usepackage{amsfonts} 
\usepackage{bm}
\usepackage[usenames,dvipsnames]{color}
\usepackage{subfigure}
\usepackage{hyperref}

\newcommand{\be}{\begin{equation}}
\newcommand{\ee}{\end{equation}}

\begin{document}

\title{Suppression of $2\pi$ phase-slip due to hidden zero modes in one dimensional topological superconductors}
\author{David Pekker$^1$, Chang-Yu Hou$^{1,2}$, Doron L. Bergman$^{1}$, Sam Goldberg$^1$, \.{I}nan\c{c} Adagideli$^{3}$, Fabian Hassler$^4$}
\affiliation{
$^1$Department of Physics, California Institute of Technology, Pasadena, CA 91125
\\
$^2$Department of Physics and Astronomy, University of California at Riverside, Riverside, CA 92521
\\
$^3$Faculty of Engineering and Natural Sciences, Sabanci University, Orhanli-Tuzla, Istanbul, Turkey
\\
$^4$Institute for Quantum Information, RWTH Aachen University, 52056 Aachen, Germany
}
\date{\today}

\begin{abstract}
We study phase slips in one-dimensional topological superconducting wires.
These wires have been proposed as building blocks for topologically protected
qubits in which the quantum information is distributed over the length of the
device and thus is immune to local sources of decoherence. However,
phase-slips are non-local events that can result in decoherence. Phase slips
in topological superconductors are peculiar for the reason that they occur in
multiples of $4\pi$ (instead of $2\pi$ in conventional superconductors). We
re-establish this fact via a beautiful analogy to the particle physics concept
of dynamic symmetry breaking by explicitly finding a ``hidden" zero mode in
the fermion spectrum computed in the background of a $2\pi$ phase-slip. Armed
with the understanding of phase-slips in topological superconductors, we
propose a simple experimental setup with which the predictions can
be tested by monitoring tunneling rate of a superconducting flux quantum
through a topological superconducting wire.
\end{abstract}

\maketitle

\section{Introduction}
\label{sec:introduction}

A quantum computer, if realized, would be able to perform computational tasks with an efficiency that could never be reached by a classical computer. Consequently, great effort has been put into exploring how to realize such a computer. One of the main challenges in doing so lies in the high sensitivity of quantum systems to background noise. Storing quantum information in topological states of matter may provide a decoherence-free realization of quantum computing. In particular, as topological states are determined by the global properties of the system, topological qubits are expected to be robust to decoherence from local perturbations~\cite{Kitaev2003}. 

We focus on a specific realization of topological matter: topological
superconducting wires. To build this type of wire one needs to combine the
properties of three discrete elements: a semiconducting nanowire that provides
strong spin orbit coupling, a superconducting wire that provides a
superconducting gap via proximity effect, and a magnetic field that opens a
Zeeman gap in the nano-wire spectrum~\cite{Sau2010, Alicea2010,Lutchyn2010,Oreg2010}. Topological superconducting wires are useful for quantum computing because a Majorana fermion forms at the interface between a conventional and a topological superconducting wire. By combining several such interfaces, it is possible to create a topological qubit as described in Ref.~\onlinecite{Alicea2011}. Further, by building a network of such wires, it is possible to perform quantum information processing by braiding the Majorana fermions, resulting in a quantum computer with topologically protected quantum logic gates~\cite{Alicea2011,Hassler2010,Sau2010a,Hassler2011,vanHeck2012}.
\begin{figure}
\includegraphics[width=8cm]{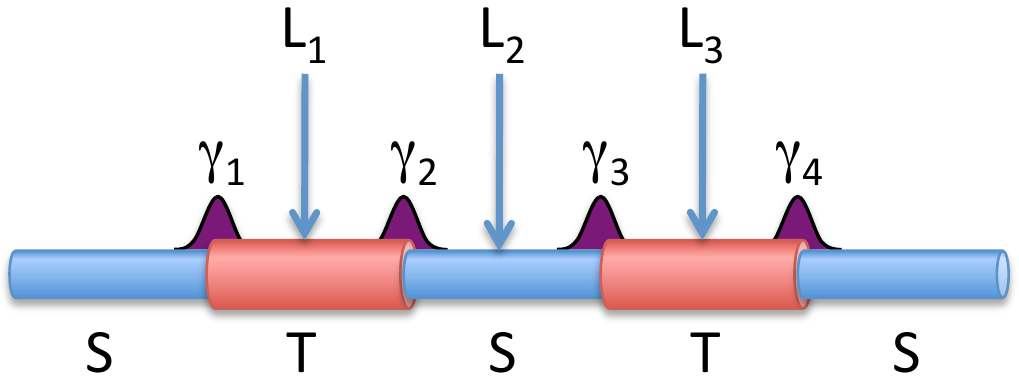}
\caption{
Schematic of a topological superconducting qubit. The qubit is composed of a series of conventional superconducting wires (labeled S) and topological superconducting wires (labeled T). Four Majorana fermions (labeled $\gamma_1$ to $\gamma_4$) are located at the interfaces. We consider phase slips at three types of locations (labeled $L_1$ to $L_3$). Topological superconductors only support $4\pi$ phase slips, which can take place at locations $L_1$ and $L_3$. These $4\pi$ phase slips do not cause decoherence of the qubit. However the central S segment (at location $L_2$) can support $2\pi$ phase slips which can cause the qubit to decohere. See main text for details.
\label{fig:qubit}
}
\end{figure}

A possible source of decoherence in such a quantum computer are phase-slips in
the superconducting wires. In a superconducting ring a phase-slip fluctuation
connects states with different winding number of the superconducting phase
around the ring. Phase-slip are fluctuations in which the amplitude of the
order parameter shrinks to zero at some location along the wire, which results
in the loss of coherence between the left and right sides of the wire, and the
phase can slip. At the conclusion of the phase slip, the order parameter
amplitude grows, and the  phase coherence is reestablished. Consequently,
phase slips play an important role in determining both the dynamics of the
order parameter as well as in determining the quantum (and the thermodynamic)
ground state of the wire. Phase-slips can be driven by either quantum or
thermal fluctuation [resulting in Quantum Phase Slips (QPS) or thermally
activated phase slips (TAPS)]. TAPS tend to dominate when the temperature is
larger than the Josephson energy for a Josephson junction, $T>E_J$ (or the
corresponding energy scale for a SC wire). In the low temperature, $T<E_J$,
thermal fluctuations become insufficient to overcome the barrier and hence QPS
become the dominant process. Experimentally, both TAPS~\cite{Webb1968,
Lukens1970, Newbower1972} and QPS~\cite{Giordano1991, Bezryadin2000, Lau2001,
Sahu2009} have been observed in thin uniform superconducting wires as well as
in constrictions~\cite{Chu2004} and Josephson junctions~\cite{Anderson1969,
Fulton1974, Martinis1987}. The effect of phase slips on topological wires has been
previously considered in Refs.~\onlinecite{Kitaev2001,vanHeck2011, Sau2011,Fidkowski2011}. The coherent motion of vortices in a 2D topological superconductor is discussed in Ref.~\onlinecite{Clarke2010,Grosfeld2011a,Grosfeld2011b}.

In this article, we investigate the effect of quantum phase slips on topological superconducting wires and devices. We start by discussing the consequence of phase slips on a superconducting qubit shown in Fig.~\ref{fig:qubit}. In particular, we note that a phase-slip of $2\pi$, which can occurs in the conventional superconducting wire segment, leads to the decoherence of the qubit while a phase slip of $4 \pi$, allowed in topological wire segments, leads to no decoherence. As phase slips can be an important source of decoherence for the topological quantum computation in Majorana fermion systems, it is important to study such processes in depth.

We consider a simplified model where phase slips only occur at a weak link (Josephson junction) in a topological superconducting wire, and construct a semiclassical field theory description for phase slips at the weak link. Then, we recover the well known fact that although the fermionic spectrum is $2\pi$ periodic in the phase difference across the weak link, the ground state has only $4\pi$ periodicity~\cite{Kitaev2001,Kwon2004a,Kwon2004b,Fu2009}. We show this in two complimentary approaches: (1) By integrating out the fermions, the partition function becomes explicitly $4\pi$ periodic. (2) We show that $2\pi$ phase slips are suppressed by making an analogy to the concept of symmetry breaking by a chiral anomaly in particle physics~\cite{Hooft1976, Rajaraman1987, Coleman1988}.

Explicitly, in method (2) we view a $2 \pi$ phase slip as an instanton event in the semiclassical description. The amplitude of the instanton is proportional to the determinant of the fermionic kernel evaluated along the instanton trajectory. Following the classic calculation of t'Hooft\cite{Hooft1976}, we explicitly obtain the eigenvalues of the fermionic kernel. We show that the spectrum contains a ``hidden" zero mode, that we uncover by a transformation of the fermionic kernel into a hermitian operator, which results in the suppression of $2\pi$ phase slips. Motivated by this result, we further discuss how the suppression of $2\pi$ phase slips can be observed by considering the effects of phase slips on topological superconductors in ring geometry (e.g. AC SQUIDs) as well as current biased topological superconducting wires.

The manuscript is organized as follows. In Sec.~\ref{sec:Phase-slips}, we discuss phase slips in a qubit device composed of topological and conventional superconducting wires. Next, we introduce the Kitaev model of a topological superconductor in Sec.~\ref{sec:setting}. We describe, in detail, QPS in topological superconducting wires and identify the hidden zero mode in Sec.~\ref{sec:weakLink}.
We discuss the detection of $4\pi$ phase slips in two types of devices made of topological superconductors: topological superconducting rings and current biased wires in Sec.~\ref{sec:devices}. Finally, we make concluding remarks in Sec.~\ref{sec:conclusions}. The main text is supplemented by two appendices, in which we derive the effective action for a topological superconducting wire with a weak link and describe the discretization of the Fermion action on the weak link in the presence of a phase-slip.

\section{Phase slips in a qubit device}
\label{sec:Phase-slips}

To motivate the study of phase slips in topological superconducting wires, we review a particular implementation of a topological qubit illustrated in Fig.~\ref{fig:qubit}. The qubit is composed of three conventional superconducting segments and two topological superconducting segments. The quantum information is stored in the four Majorana states labeled $\gamma_1$ to $\gamma_4$. To describe how quantum information is stored we use the basis of complex fermions $c_L=\gamma_1+i\gamma_2$ and $c_R=\gamma_3+i\gamma_4$, for the ``left'' and ``right'' topological segments. We can describe the state of the device in terms of the occupation numbers $|n_L,n_R\rangle$ of the left and right complex fermions. For states of odd parity, we could use $|0,1\rangle$ and $|1,0\rangle$ to represent the two states of the qubit. Analogously, for states of even parity, we could used $|0,0\rangle$ and $|1,1\rangle$ to represent the two states of the qubit.

Consider the effect of phase slips on the qubit device illustrated in Fig.~\ref{fig:qubit}. Phase slips that can potentially damage the quantum information in the qubit can occur at three typical locations labeled $L_1$, $L_2$, and $L_3$. Locations $L_1$ and $L_3$ lie inside topological superconducting segments and, as we shall show later, only support $4\pi$ phase slips only. On the other hand, $L_2$ lies inside a conventional superconductor and thus can supports $2 \pi$ phase slips.

To understand how a phase slip can affect the quantum information stored in a qubit, we appeal to the Aharonov-Casher effect~\cite{Aharonov1984}. A $2\pi$ phase slip can be thought of as taking a vortex on a closed loop trajectory around the wire, with the trajectory intersecting the wire at the location of the phase slip. The Aharonov-Casher effect states that when we take a flux around a charge on a closed trajectory, the wave function builds up a phase proportional to the charge enclosed. In particular, when a single superconducting vortex goes once around a single electron charge, the sign of the wave function changes. 

\begin{figure}
\includegraphics[width=8cm]{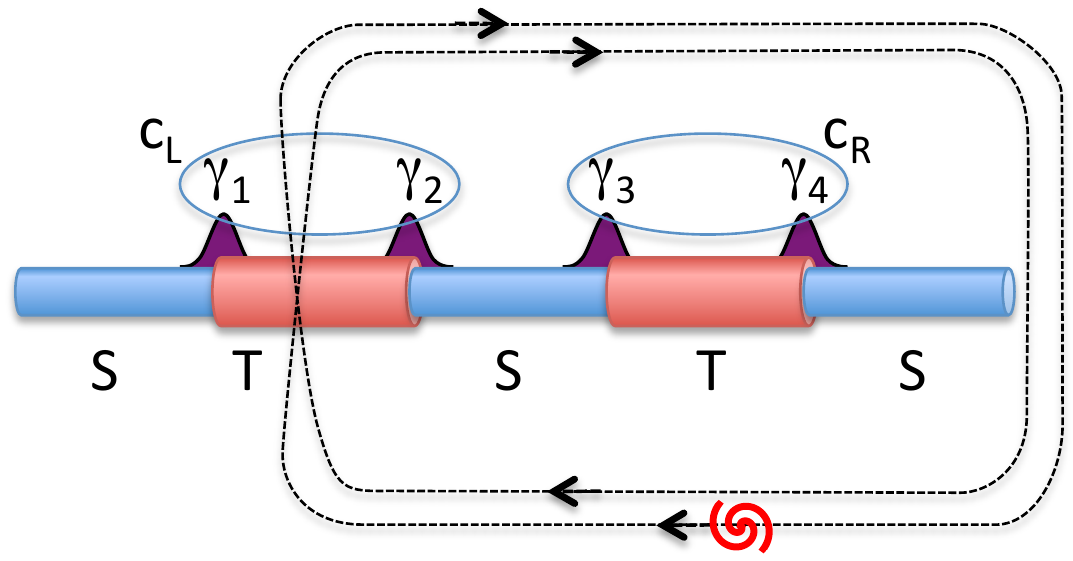}
\caption{
Schematic of a vortex trajectory equivalent to a $4\pi$ phase slip at location $L_1$ of Fig.~\ref{fig:qubit}.
\label{fig:trajectory}
}
\end{figure}

Let us first consider a phase slip at location $L_1$ as depicted in Fig.~\ref{fig:qubit}. Since $L_1$ lies inside a topological superconductor, only $4\pi$ phase slips are supported, which is equivalent to a vortex completely encircling the right segment of the topological superconducting wire twice as depicted in Fig.~\ref{fig:trajectory}. The double encirclement means that the phase of the wave function associated with fully encircled fermions is unchanged irrespective of their occupation numbers according to the Aharonov-Casher effect. Hence, there is no overall phase accumulation related to the occupation number of the $c_R$ fermion in Fig.~\ref{fig:trajectory}.

On the other hand, the effect of a $4 \pi$ phase slip on the quantum state when the vortex core crosses through a delocalized fermion, as is the case for $c_L$ fermion in Fig.~\ref{fig:trajectory}, is more delicate. To work out this scenario, we consider a special setting where the phase slip occurs at a weak link. In the limiting case of an extremely weak link of the topological superconducting wire, there will be a localized fermion $c_w \equiv \gamma_{w,L}+i\gamma_{w,R}$ associated with the weak link. Here, $\gamma_{w,L(R)}$ are Majorana fermions residing at the left (right) of the weak link and can be combined with the constituent Majoranas of the $c_L$ fermion to form two fermions $c_{L1}$ and $c_{L2}$ that are localized to the left and to the right of the weak link. By the Aharonov-Casher effect, the wave function of $c_{L1}$  and $c_{L2}$ fermions returns to its initial value following a $4\pi$ phase slip. Therefore, the $c_L$ fermion also returns to its initial state~\cite{Kitaev2001}. We shall give explicit arguments on how this occurs for the generic case in appendix~\ref{app:weakLink}.

Combining the results of the previous two paragraphs, we conclude that a $4\pi$ phase-slip at $L_1$ brings the qubit back to its initial quantum state and does not cause decoherence. In a similar manner, one can argue that a $4 \pi$ phase slip at $L_3$ does not change the quantum state. The only difference is that the vortex encircles the ``inactive" $c_R$ fermion twice for a $4\pi$ phase-slip at $L_1$, which accumulates a phase of $2\pi$, while it does not encircle the inactive fermion $c_L$ for the phase-slip at $L_1$, which brings no extra phase.

Finally, we consider the effect of a $2\pi$ phase slip at position $L_2$. Again, we let the phase to the right of the phase-slip core wind by $2\pi$ while the phase to the left remains unchanged. Here we find that states with $c_R$ fermion empty remain unchanged ($|1,0\rangle\rightarrow|1,0\rangle$, $|0,0\rangle\rightarrow|0,0\rangle$), while those with $c_R$ occupied acquire a minus sign ($|0,1\rangle\rightarrow-|0,1\rangle$, $|1,1\rangle\rightarrow-|1,1\rangle$). Therefore, phase-slips at $L_2$ decohere the qubit~\footnote{We have assumed that there are even number of fermions to the right of the phase slip center, not counting the $c_R$ fermion. The consequences for the minus sign are reversed for the case with an odd number.}. 

\section{Setting: topological superconducting wires}
\label{sec:setting}

To make concrete arguments about phase slips in topological superconducting wires, and devices containing topological topological wires, we focus on the implementation of topological superconducting wires described in Ref.~\onlinecite{Alicea2010}. In this implementation, topological superconductivity is not obtained as an intrinsic property of a material, but rather by combining various materials to engineer the desired properties. The main part of the proposed composite is a single channel semiconducting nanowire with strong spin orbit coupling. By applying a strong magnetic field, the electrons in the nanowire form two, well separated, spin polarized bands. Due to the presence of both a magnetic field and the spin orbit scattering, the spin polarization in the two bands is momentum dependent. Finally, by proximity coupling the semiconducting nanowire to a conventional s-wave superconductor, we induce p-wave pairing in the bottom band of the nanowire. Thus the semiconducting nanowire is predicted to exhibit topological superconductivity. 

\begin{figure}
\includegraphics[width=8cm]{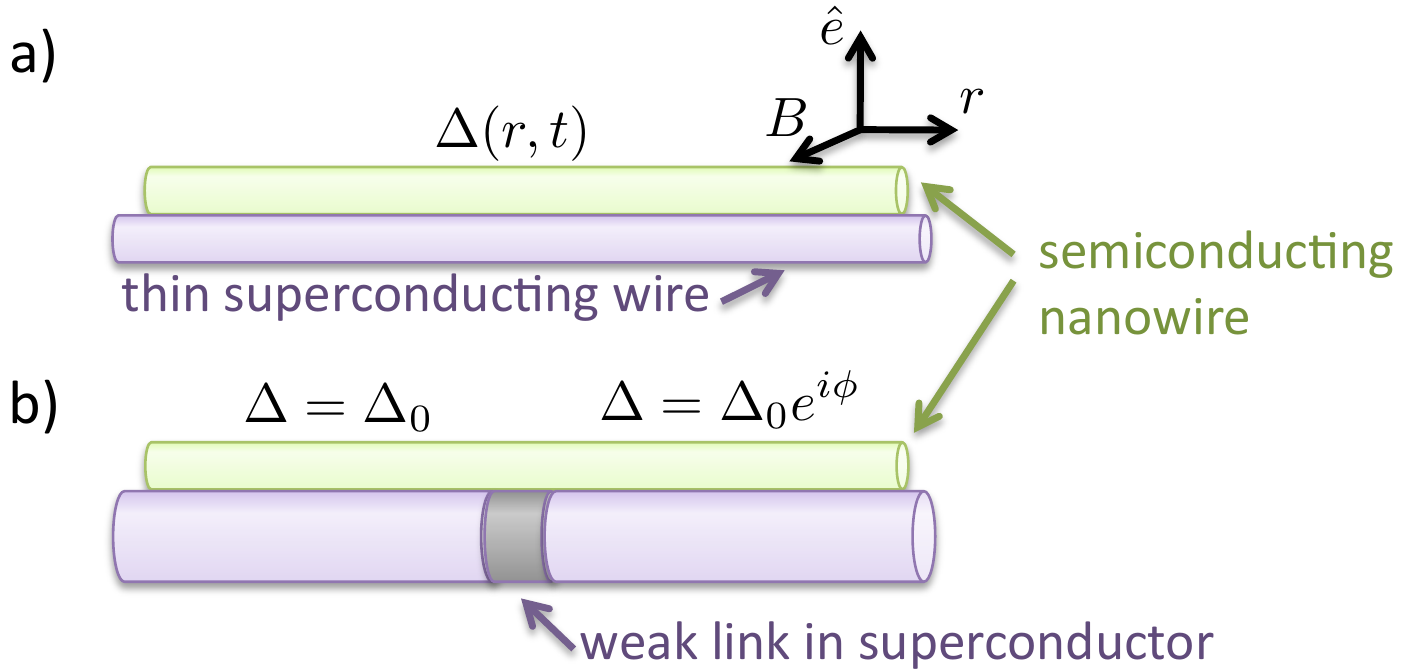}
\caption{
Schematic of a composite structure consisting of a semiconducting nanowire in contact with a  superconductor. The superconductor induces pairing in the nanowire via proximity effect. The orthogonal alignment of the spin-orbit field $\hat{e}$, the magnetic field $B$, and the coordinate along the wire $r$ is indicated. In implementation (a) the superconductor is a thin homogenous wire that is susceptible to phase slips along it's entire length. In implementation (b) the superconductor is rigid everywhere except a weak link, a point at which phase-slips can occur.
\label{fig:composite}
}
\end{figure}

In the composite implementation of topological superconductivity, phase slips in the topological superconductor are associated with phase slips in the proximity giving superconductor. We therefore assume that the superconductor is sufficiently weak so that it can support phase-slip fluctuations. This can occur if the superconductor is a sufficiently narrow wire~\cite{Giordano1991, Bezryadin2000, Lau2001}, or if there is a weak spot or break in the superconductor which results in the formation of a Josephson junction. These possibilities are schematically illustrated in Fig.~\ref{fig:composite}. To model the composite structure, we use the Kitaev model~\cite{Kitaev2001} to describe the electrons in the semiconducting nanowire, and supplement it with a phenomenological model that describes the order parameter in the proximity giving superconductor. 

The Kitaev model is specified by the Hamiltonian 
\begin{align}\label{H0}
\hat H=-\sum_{i=1}^N \mu_i c_i^\dagger c_i -\sum_{i=1}^{N-1}[tc_{i+1}^\dagger c_i + \Delta_{i,i+1} c_{i+1}^\dagger c_i^\dagger+h.c.] 
\end{align}
where $N$ is the number of lattice sites, $c_i^\dagger$ ($c_i$) is the
electron creation (annihilation) operator at site $i$, $\mu_i$ is the chemical
potential at site $i$, $t>0$ is the hopping matrix element, and $\Delta_{i,i+1}$
is the complex order parameter, defined on the link between sites $i,i+1$.
This model can be thought of as the large magnetic field regime of the
model described in Ref.~\onlinecite{Alicea2010}. The model supports both
topological and conventional phases by tuning of the chemical potential, with
the phase transitions occurring at $|\mu|=2t$. Thus, we can model both topological and conventional segments by varying $\mu_i$ as a function of position along the wire. 

To describe the dynamics of the order parameter in the superconductor, we need to choose whether we are describing a Josephson junction or a continuous thin superconducting wire. As we are interested in the effect of the electron degrees of freedom on phase slips, these details will not be especially important. In the next section, we shall focus on the technically simpler problem of phase slips at a Josephson junction (weak link). 

\section{Phase slips at a weak link: the hidden zero mode}
\label{sec:weakLink}

In this section, we construct a theory of phase slips in the weak link
geometry illustrated in Fig.~\ref{fig:composite}(b): a semiconducting wire on
top of a superconducting wire with a single weak link. We start with this
geometry as it involves fewer degrees of freedom than the continuous wire
geometry illustrated in Fig.~\ref{fig:composite}(a). 

We explicitly construct an effective, low energy, model of the weak link geometry starting from the Kitaev model~\eqref{H0} in appendix~\ref{app:weakLink}. From the point of view of superconductivity, the weak link geometry is a Josephson junction, that can be characterized by the phase difference $\phi$ across the weak link. From the point of view of the electrons in the semiconducting nano-wire, the weak link is a topological-conventional-topological junction. Associated with each topological-conventional interface, there is a Majorana fermion.  By assumption, the weak link is short compared to the Fermi-wavelength in the nanowire, and therefore the two Majorana fermions interact to form a single complex fermion $c_{w}$ that is localized on the weak link. The low frequency effective action involves $\phi$ and $c_w$ degrees of freedom associated with the weak link and is given by
\begin{align}
\label{eq:Seff}
S_\text{J}
= \int d t \left[ \frac{1}{2}\frac{1}{8 E_C} \left(\partial_t \phi\right)^2 - E_J (1- \cos(\phi)) \right. 
\\
+
\left. c^{\dag}_w \Big( i \partial_t-E_M \cos(\phi/2) \Big) c_w \right] .\nonumber
\end{align}
In this model, the first term is phenomenological in origin and describes the charging energy $E_C=e^2/2C$ due to the capacitance $C$ associated with the weak link. The $E_J$ term describes the $2\pi$ periodic part of the potential energy and is primarily related to the electronic states of the semiconducting nanowire outside the gap. There can be a secondary contribution to the $E_J$ term from the Josephson energy associated with the weak link in the underlying superconductor. The final term describes the sub-gap fermion $c_w$, localized at the weak link. The energy scale $E_M$ and $E_J$ can be obtained from the Kitaev model, see appendix~\ref{app:weakLink}. 

\begin{figure}
\includegraphics[width=7cm]{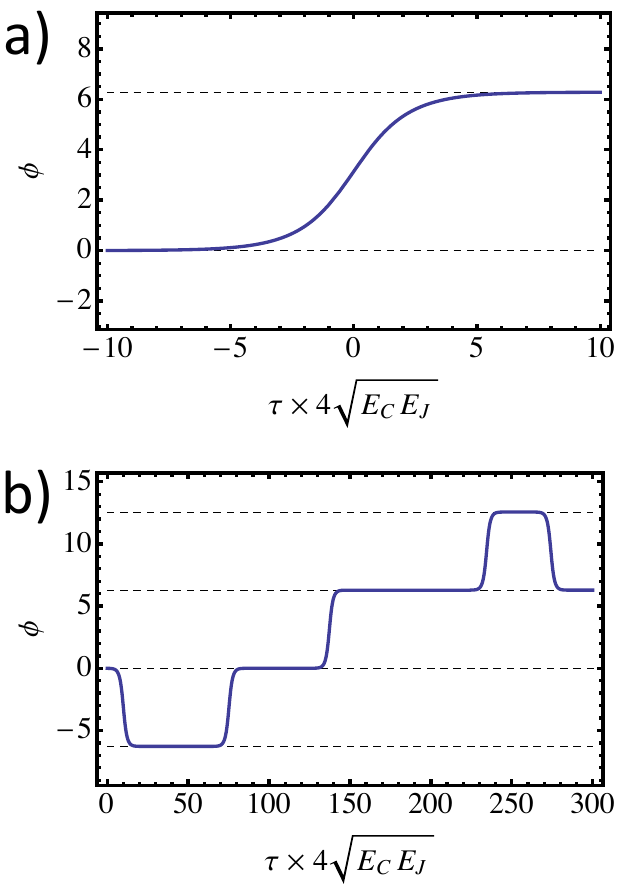}
\caption{
(a) Instanton trajectory in the sine-Gordon model, Eq.~\eqref{eq:sGE}. (b) Schematic representation of the dilute instanton gas composed of $2\pi$ phase slips and $-2\pi$ anti-phase-slips.
\label{fig:instanton}
}
\end{figure}

We begin by sketching the semi-classical dynamics of the phase only (i.e. sine-Gordon) model, without the fermionic term, as described by the real time action
\begin{align}
S_\phi
=\int dt\, \left[ \frac{1}{2}\frac{1}{8 E_C} \left(\partial_t \phi\right)^2 - E_J\left(1-\cos(\phi)\right) \right].
\end{align}
The potential energy associated with the second term of this action is $2\pi$ periodic, thus we would expect that the phase would be localized near $0, \pm 2\pi, \pm 4\pi \dots$. However, quantum fluctuations driven by the first term can connect these minima via phase slips. Following the instanton prescription, we can obtain a semiclassical approximation for the tunneling matrix element~\cite{Rajaraman1987, Coleman1988}. The prescription states that we must first go to the imaginary time (Euclidean) description via $t\rightarrow i \tau$
\begin{align}
\tilde{S}_\phi
=\int d\tau\, \left[ \frac{1}{2}\frac{1}{8 E_C} \left(\partial_{\tau} \phi\right)^2 + E_J \left(1-\cos(\phi)\right) \right].
\label{eq:sGE}
\end{align}
Going to the Euclidean description results in the change of the sign of the
potential energy term. Thus, the minima at $0$ and $2\pi$ in the real time
description, become maxima in the Euclidean description. Moreover, in the
Euclidean description there is a classical trajectory $\phi_\text{cl}(\tau)$
that connects these maxima: $\phi_\text{cl}(-\infty)=0$ and
$\phi_\text{cl}(\infty)=2\pi$, which is illustrated in
Fig.~\ref{fig:instanton}(a). The instanton trajectory leads to the value of the tunneling matrix element, which at lowest order is
\begin{align}
\langle 0 | e^{i H t} | 2\pi \rangle \sim e^{-\tilde{S}_\phi[\phi_\text{cl}]}.
\end{align}
where $\tilde{S}_\phi[\phi_\text{cl}]$, is the value of the action associated with the classical trajectory $\phi_\text{cl}(\tau)$. 

A complimentary approach to studying dynamics is to study the thermodynamical ground state. Instanton trajectories extremize the action and are therefore important in the description of the thermodynamic ground state. Indeed, we can think of the low temperature ground state, associated with $\tilde{S}_{\phi}$, as a dilute gas of phase slips and anti-phase-slips~\cite{Rajaraman1987,Coleman1988}, which is schematically illustrated in Fig.~\ref{fig:instanton}(b). 

At this point we are ready to ask the question of what is the effect of Fermions, i.e. the third term in Eq.~\ref{eq:Seff}, on the phase slips and therefore on the ground state. To answer this question, we consider the partition function corresponding to the thermodynamic ground state
\begin{align}
\label{eq:partition-function-full}
Z=\int {\cal D} \phi \, {\cal D} c_w\, {\cal D}c_w^\dagger\, e^{-{\tilde{S}_\text{J}}},
\end{align}
where $\tilde{S}_\text{J}$ is the Euclidean action associated with $S_\text{J}$. We are particularly interested in the low temperature regime $T\rightarrow0$, in which the integral in $\tilde{S}_\text{J}$ runs over a long
stretch of imaginary time from $\tau=0$ to $\tau=\beta=1/T$. We will answer the question about the role of the fermions in two ways. First, we will integrate out the fermions and obtain an effective phase-only partition function that takes into account the contribution of the fermions. Second, we will appeal to a beautiful analogy to a problem in particle physics to show how the fermionic term breaks $2\pi$ phase rotation symmetry in the ground state.

\subsection{Method 1: Integrating out fermion}
\label{subsec:m1}

In this subsection, our goal is to integrate over the fermionic degrees of freedom in the partition function and convert the action Eq.~\eqref{eq:Seff} to an effective action depending only on the phase, $\phi$. Since the fermionic part of the Lagrangian is quadratic, we can integrate over the fermionic degrees of freedom in Eq.~\eqref{eq:partition-function-full} for an arbitrary trajectory $\phi(\tau)$ and obtain the expression
\begin{align}
Z \propto \int {\cal D} \phi \det[{\mathcal K}_f(\phi)]e^{-\tilde{S_\phi}}.
\label{eq:Zf}
\end{align}
Here, we use the proportionality sign to accommodate the normalization of the fermion path integral, and ${\mathcal K}_f(\phi)$ is the Lagrangian density of the fermionic part of the action
\begin{equation}
\label{eq:Kf}
\begin{split}
S_f(\phi) =& \int d\tau \, c_w^\dagger {\mathcal K}_f(\phi) c_w , 
\\
=& \int d\tau c_w^\dagger \left[ \partial_\tau+E_M \cos(\phi/2) \right] c_w, 
\end{split}
\end{equation}
subject to an anti-periodic boundary condition $c_w(\beta)=-c_w(0)$.

To compute the fermionic determinant, we make use of the fact that $\det[{\mathcal K}_f(\phi)] = \prod_n \lambda_n$, where $\lambda_n$'s come from the eigenvalue problem
\begin{equation}
\label{eq:Kf-diff-eq}
{\mathcal K}_f(\phi)
u_n(\tau)
=\lambda_n
u_n(\tau).
\end{equation}
Solving the eigenvalue problem, for arbitrary $\phi(\tau)$, we find the implicit expression for the eigenfunctions $u_n$
\begin{equation}
\label{eq:un-K}
u_n(\tau)=e^{\int_0^\tau \left( \lambda_n - E_M \cos[\phi(\tau')/2] \right) \,d\tau'}.
\end{equation}
With the anti-periodic boundary conditions, we obtain
\begin{align}
\label{eq:spectrum-anti}
 \lambda_n= \frac{i \pi (2 n+1)}{\beta} + \frac{I_1}{\beta} 
\end{align}
where $I_1=\int_0^\beta d\tau\,E_M \cos(\phi/2)$ and $n$ is an integer.
Using a few well known identities as in Ref.~\onlinecite{Dashen1975}, we now find
\begin{equation}
\label{eq:det-Kf}
\begin{split}
\det[{\mathcal K}_f(\phi)] =&\prod_n\left(\frac{i \pi (2 n+1)}{\beta} + \frac{I_1}{\beta}
\right),
\\
=& \left[\prod_n\left(\frac{i \pi (2 n+1)}{\beta}\right)\right]
\cosh(I_1/2). 
\end{split}
\end{equation}
Thus we find that the partition function becomes
\begin{align}
 Z \propto
  Z_\text{eff}=\int {\cal D} \phi\, \cosh(I_1/2) e^{-\tilde{S}_\phi}.
\end{align}
We interpret this partition function as follows. The fermion can be in one of two states (either even or odd parity), since there are no terms in the Hamiltonian that connect these states, the partition function splits into two parts: one part for even parity and the other part for odd parity, manifested in 
$\cosh(I_1/2) e^{-\tilde{S}_\phi} = \frac{1}{2} \left[ e^{-\tilde{S}_\phi - I_1/2} +
e^{-\tilde{S}_\phi + I_1/2} \right]$. The even and odd parity states are separated by the energy $E_M \cos(\phi/2)$, and the effective action becomes 
\begin{align}
\tilde{S}_{\phi-\text{eff}}=\int_0^\tau \left[ d\tau \, \frac{1}{2}\frac{1}{8 E_C} \left(\partial_t \phi\right)^2 - E_J (1- \cos(\phi))\right. \nonumber\\
\left. \pm \frac{E_M}{2} \cos(\phi/2)\right],
\end{align} 
where the sign of the last term is determined by the parity of the fermionic state. We note that this action is called the double sine-Gordon model.

\subsection{Method 2: the hidden zero mode}
\label{subsec:m2}

As we have argued, the quantum (as well as the low temperature thermodynamic)
ground state of $\tilde{S}_\phi$ is composed of a superposition of quantum states where $\phi$ localized at multiples of $2\pi$. Due to quantum fluctuations, states with different $\phi$'s are connected by instantons. In this subsection, we explicitly show that this picture is significantly modified in the presence of the fermion degree of freedom, by considering the fermionic path integral in the background of a phase slip. Indeed, what we find is that $2\pi$ phase slips are strongly suppressed by the appearance of a ``hidden" zero mode in the fermionic determinant. As a result, the $2\pi$ periodic symmetry of the spectrum is broken down to $4\pi$ periodic symmetry in accord to the effective action that was obtained in the previous section. This mechanism of symmetry breaking was first studied in the context of high energy physics, specifically it was used by t'Hooft to explain the ``missing meson" problem of quantum chromodynamics in Ref.~\onlinecite{Hooft1976}, see also Refs.~\onlinecite{Rajaraman1987, Coleman1988}. 

Consider a bounce (phase-slip followed by an anti-phase-slip), such that $\phi(0)=2\pi$ and $\phi(\beta)=0$. To be concrete, we will focus on phase slips with the functional form $\cos(\phi(\tau)/2)=\tanh \left(\frac{\tau-\beta/2}{w}\right)$.
In describing the rare instanton gas, the instantons must be separated by long stretches of imaginary time. Therefore, to understand a single instanton, we must look towards the limit $\beta \rightarrow \infty$. What do we expect in this regime? Following the above discussion of the partition function, we expect that the matrix element must be (see Ref.~\onlinecite{Rajaraman1987, Coleman1988} for details)
\begin{align}
\langle 0 | e^{i H t} | 2\pi \rangle \propto \frac{\det[{\mathcal K}_{f,2\pi}] }{\det{[{\mathcal K}_{f,0}]}} e^{-\tilde{S}_\phi[\phi_\text{cl}]},
\end{align}
where ${\mathcal K}_{f,2\pi}$ (${\mathcal K}_{f,0}$) is the Lagrangian density
operator in the presence (absence) of a bounce. ${\mathcal K}_{f,0}$ is
necessary for normalization. In what follows, we will use the similar subscripts $_{2\pi} (\,_0)$ to indicate operators in the presence (absence) of a bounce. Specifically using
Eq.~\ref{eq:det-Kf}, the ratio of fermion determinants is
\begin{align}
\frac{ \det[{\mathcal K}_{f,2\pi}]}{ \det{[{\mathcal K}_{f,0}]}}=\frac{\cosh\left(\frac{1}{2}\int_0^\beta \cos(\phi[\tau]/2) \, d\tau\right)}{\cosh(\beta/2)}, \label{k2}
\end{align}
which becomes $\sim e^{-\beta/2}$ in the limit $\beta \rightarrow \infty$,
since with the bounce, the integrand in the numerator will be negative for a large part of the 
interval $\left[0,\beta\right]$, and thus $\int_0^\beta \cos(\phi[\tau]/2) \, d\tau \ll \beta$.

To uncover the ``hidden" zero mode in the fermionic determinant, we first rewrite the fermionic action, Eq.~\eqref{eq:Kf}, in a doubled form
\begin{align}
\label{eq:Lf}
&S_f(\phi) = \int d\tau \, \psi^\dagger {\mathcal L}_f(\phi) \psi
\\
&= \int d\tau \psi^\dagger \left(
\begin{array}{cc}
\partial_\tau+E_M \cos(\phi/2) & 0\\
0 & \partial_\tau-E_M \cos(\phi/2)
\end{array}\right) \psi, \nonumber
\end{align}
where $\psi^{\dagger} =\left( c^\dagger_w, c_w \right)$, subjected to anti-periodic boundary conditions $\psi(\beta)=-\psi(0)$. Evidently, we have $\det[{\mathcal K}_f(\phi)] = \sqrt{\det[{\mathcal L}_f(\phi)]}$, which can be shown explicitly by using the fact $\text{det}[\mathcal{L}_f] =  \prod_i \bar{\lambda}_i$, where $\bar{\lambda}_i$ are eigenvalues of the differential equations
\begin{align}
\label{eq:Lf-diff-eq}
{\mathcal L}_f(\phi)
\left(\begin{array}{c}
u_i(\tau)\\
v_i(\tau)
\end{array}\right) 
=\bar{\lambda}_i
\left(\begin{array}{c}
u_i(\tau)\\
v_i(\tau)
\end{array}\right). 
\end{align}
The eigenvalues $\bar{\lambda}_i$ can be obtained in the similar way as Eqs.~\eqref{eq:Kf-diff-eq},~\eqref{eq:un-K} and~\eqref{eq:spectrum-anti}, and take the form  $\bar{\lambda}_n^{\pm}= \frac{i \pi (2 n+1)}{\beta} \pm \frac{I_1}{\beta}$ for all integer $n$. Here, $\bar{\lambda}_n^+$ correspond to $u_i$ sector while $\bar{\lambda}_n^-$ correspond to $v_i$ sector. As expected, the product of all $\bar{\lambda}_{i}$ gives $\text{det}[\mathcal{K}_f(\phi)]^2$.

To facilitate the analysis, we transform the differential operator $\mathcal{L}_f$ in Eq.~\eqref{eq:Lf-diff-eq} into a difference operator $L_f$. By discretizing the interval $\tau \in [0,\beta]$ with $N$ lattice points, we first arrange the amplitudes of the wave function at each lattice site, $u_{n}$ and $v_{n}$ with $n \in 1,\dots, N$, in a vector form
\begin{equation}
\Xi= (u_1,u_2,\dots ,u_N, v_1,v_2, \dots ,v_N)^{T}, 
\end{equation}
Then, the difference equation corresponding to Eq.~\eqref{eq:Lf-diff-eq} becomes $L_f \Xi =\lambda \Xi $, where the difference operator takes the form $L_f = L_{f}^{u}\oplus L_{f}^{v}$. We then have
\begin{equation}
\label{eq:Lf-uv}
\begin{split}
 L_{f}^{u}=&\left[ \frac{1}{2\delta} \left(\delta_{i+1,j}-\delta_{i,j+1}  \right) + \Delta_{i} \delta_{i,j} \right],
\\
L_{f}^{v}=& \left[ \frac{1}{2\delta} \left(\delta_{i+1,j}-\delta_{i,j+1}  \right) -\Delta_{i} \delta_{i,j} \right],
\end{split}
\end{equation}
where $i,j \in 1,\dots,N$, $\Delta_n=\cos(\phi(n \delta)/2)$ and $\delta=\beta/N$ is the step in imaginary time. Now, the determinant of the difference operator $\mathrm{det}[L_f]$ is simply the product of all eigenvalues of $\lambda$.

However, discretization scheme in Eq.~\eqref{eq:Lf-uv} suffers from the notorious fermion doubling problem and effectively doubles the number of fermions both for $u(\tau)$ and $v(\tau)$ sectors~\cite{Negele1998}. Hence, the continuum limit of the determinant $\mathrm{det}[L_f]|_{N \to \infty}$ is not associated with $\mathrm{det}[\mathcal{L}_f]$ directly. Instead, one expects the relation $\mathrm{det}[L_f]|_{N \to \infty} \sim \mathrm{det}[\mathcal{L}_f]^2$. By introducing the proper normalization as in Eq.~\eqref{k2}, we find
\begin{equation}
\left.\frac{\det[ L_{f,2\pi}]}{\det[ L_{f,0}]}\right|_{N\to \infty}
= 
\frac{\det[{\mathcal L}_{f,2\pi}]^2}{\det[{\mathcal L}_{f,0}]^2}
=
\frac{ \det[{\mathcal K}_{f,2\pi}]^4 }{ \det[{\mathcal K}_{f,0}]^4 } .
\end{equation}

\begin{figure}
\includegraphics[width=8cm]{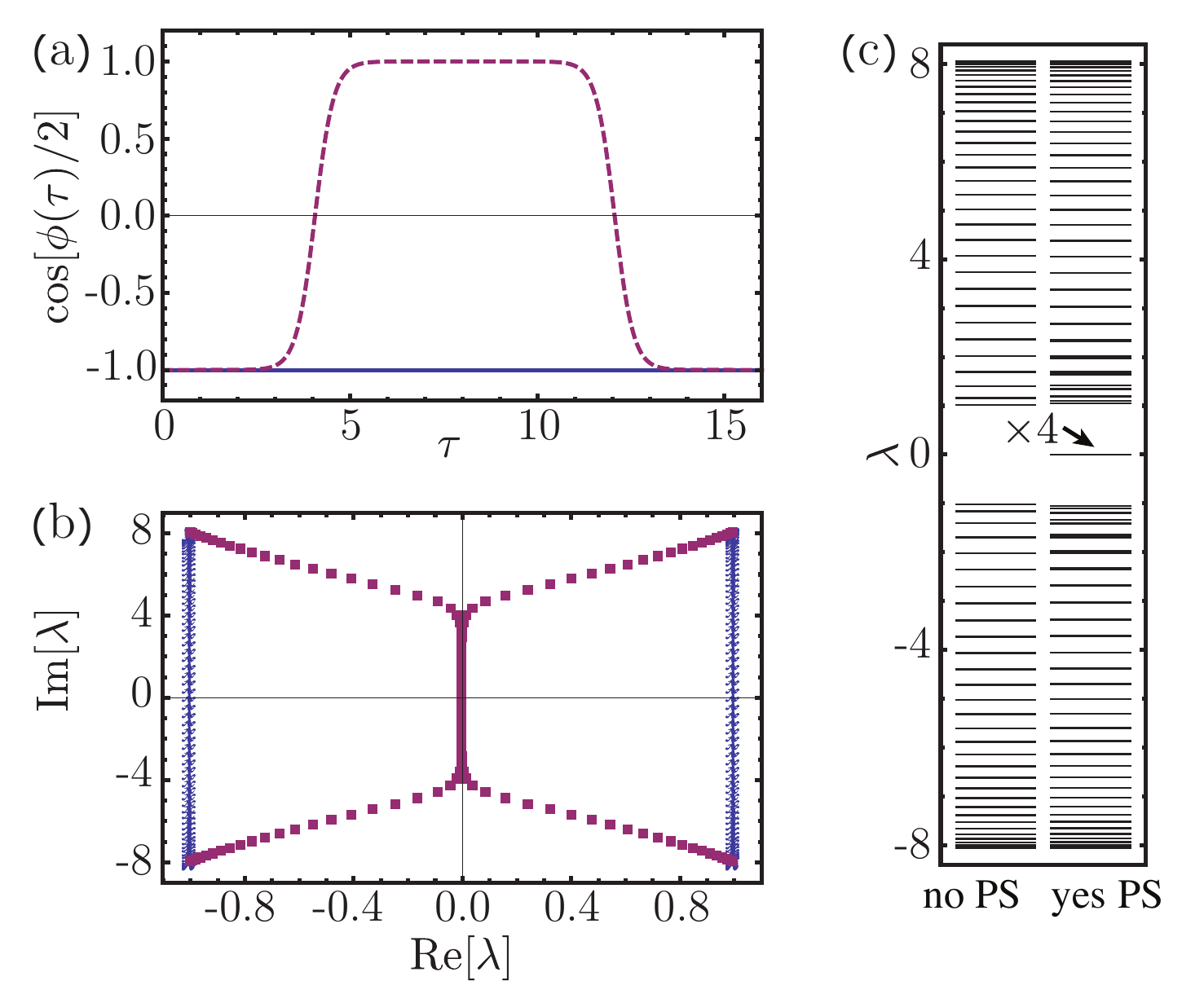}
\caption{
(a) $\cos[\phi(\tau)/2]$ as a function of $\tau$ for the no phase-slip case (blue), and a phase-slip followed by an anti-phase-slip trajectory (red), using $\beta=16$. 
(b) Eigenspectrum of ${L}_{f,2\pi}$ with antiperiodic boundary conditions, $\beta=16$, $n_\tau=128$. Blue dots represent the spectrum with no phase slips and red dots represent the spectrum with a phase-slip followed by an anti-phase slip.
(c) Eigenspectrum of $T \cdot {L}_{f,2\pi}$, no phase slip on the left and phase-slip followed by an anti-phase slip on the right. The fermionic spectrum on the right contains four zero modes. 
}
\label{fig:spectrum}
\end{figure}

We compute the spectrum of the difference operator ${L}_{f}$ using
anti-periodic boundary conditions with constant $\phi(\tau)$ and with a $2\pi$
phase slip followed by a $2\pi$ anti-phase-slip, see
Fig.~\ref{fig:spectrum}(a). We have to use a phase-slip followed by an
anti-phase-slip in order to make the boundary conditions on the fermions make
sense. Without phase-slips, the eigenspectrum of ${L}_{f,0}$ contains two
lines of eigenvalues in the complex plane with $\text{Re} \lambda_i=\pm E_M$,
see Fig.~\ref{fig:spectrum}(b). In the presence of the phase slips, the
eigenspectrum deforms as plotted in Fig.~\ref{fig:spectrum}(b). However, in the presence of phase-slips, the spectrum contains no obvious zero modes. 

The final step needed to uncover the zero mode is to consider the operator $H_f = T\cdot L_f$, where
$T=i \sigma^y \otimes \openone_{N}$ and $\openone_N$ is a $N\times N$ identity
matrix. We note that this transformation does not change the determinant,
$\det [H_{f}] =  \det [L_{f}]$ (up to a sign, which gets cancelled in the
normalization). While the operator $L_{f}$ is not hermitian, the
transformed operator $H_f$ is hermitian. Indeed, the eigenspectrum of the
$H_f$ operator without phase slips looks like a gapped spectrum, with the gap
set by $E_M$, see Fig.~\ref{fig:spectrum}(c). On the other hand, for the
phase-slip followed by an anti-phase-slip $\phi(\tau)$ trajectory depicted in
Fig.~\ref{fig:spectrum}(a), we find that the gap is occupied by four modes with near zero eigenvalues. As the splitting of these modes from zero depends exponentially on the separation of the two phase-slips, we shall refer to these modes as the zero modes.

On closer inspection, the $H_f$ Hamiltonian looks like the Hamiltonian of polyacetylene. In continuum notation, the operator $H_{f}$ is
\begin{align}
H_{f}=\left(
\begin{array}{cc}
0 & -\partial_\tau-E_M \cos(\phi/2)\\
\partial_\tau-E_M \cos(\phi/2) & 0
\end{array}\right)
\end{align}
where $\tau$ represents the position along the polyacetylene chain.  Now, we
can leverage the well known properties of the polyacetylene Hamiltonian to
understand our Josephson junctions action: each time the mass changes sign
(i.e. $\phi$ phase slips by $2 \pi$) there appears an extra zero mode that is
localized on the kink (phase-slip). Because of Fermion doubling, in the
discrete version we actually find two zero-modes associated with each kink. In
case there is more than one kink, the zero modes will be split, with the
splitting being exponentially suppressed in the separation of the kinks.
Indeed, in Fig.~\ref{fig:spectrum}(c) we see a signature of this effect, with four zero modes appearing in the gap, once we introduce two kinks (a phase-slip followed by an anti-phase-slip). In summary, going back to the original undoubled model Eq.~\eqref{eq:Seff}, each phase-slip is associated with $1/2$ zero mode. 

We pause to remark on the relation between the boundary conditions and the
zero modes. In principle, we can choose open, periodic, anti-periodic or some
other form of boundary conditions. Despite the choice of boundary conditions,
each $2\pi$ phase slip will result in the appearance of two additional zero
modes in the discretized model. We note that for the case of anti-periodic (or
periodic) boundary conditions, in order for the sign of $E_M
\cos(\phi(\tau)/2)$ to match across the boundary, phase slips must be added in
multiples of $4\pi$. Finally, we add that in order to obtain the correct value
of the partition function, we must indeed use anti-periodic boundary
conditions, see appendix of Ref.~\onlinecite{Dashen1975}.

\begin{figure}
\includegraphics[width=8cm]{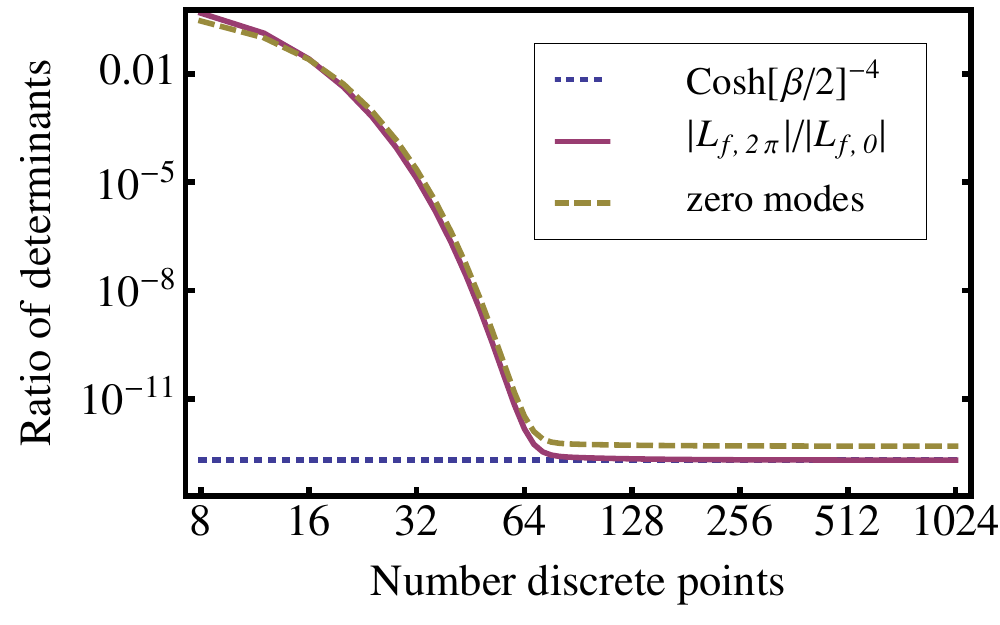} 
\caption{
Ratio of determinants for the phase profile pictured in
Fig.~\ref{fig:spectrum}(a) computed using different methods. (1) following the prescription of Method 1 we integrate out the fermionic degrees of freedom without discretization, and raise the final answer to the fourth power to compensate for the two fermion doublings in the discretized methods (labeled: $\cosh(\beta/2)^{-4}$). (2) following prescription of Method 2, we compute the fermion determinants on a discrete lattice (labeled: $|L_{f,2\pi}|/|L_{f,0}|$). (3) following Method 2, by constructing the ratio of the four smallest fermion eigenvalues $(\lambda_{1,2\pi}/\lambda_{1,0})\times \dots \times (\lambda_{4,2\pi}/\lambda_{4,0})$ of $H_{f,2\pi}$ and $H_{f,0}$, respectively (labeled: zero modes). Comparison of the three curves indicates that the suppression of tunneling is indeed controlled by the zero modes, with the small offset being a non-universal feature associated with the duration of the phase slip. 
}
\label{fig:det}
\end{figure}

Having found that phase-slips in the order parameter are associated with
zero-modes in the fermion determinant, we now demonstrate that these zero
modes indeed control the value of the fermion determinant. To test this, we
consider the two trajectory depicted in Fig.~\ref{fig:spectrum}(a). First, as a consistency check, we compute the ratio of determinants for this pair of trajectories using both the continuum method described in the previous subsection and the discrete method described in this subsection. To make a direct comparison, we square the continuum result in order to match the effects of fermion doubling. We plot the comparison, as a function of the number of discretization steps in Fig.~\ref{fig:det}. The figure demonstrates that the two ways to compute the ratio of the fermion determinants converges as the number of discretization steps increases. Next, we compare the ratio of the determinants to the ratio of the four smallest eigenvalues, i.e. the product of four eigenvalues of near zero modes divided by the quartic of the gap, $(E_M/2)^4$. We see that the ratio of the eigenvalues follows closely the ratio of the determinants computed using the discrete method, except for a small offset of order unity, see Fig.~\ref{fig:det}. The offset is associated with the imaginary time size of the phase slip. Thus the ratio of determinants is indeed controlled by the zero modes. 

In appendix~\ref{app:disc}, we shall describe an alternative discretization
scheme for avoiding the fermion doubling with the cost that the spectrum of
the difference operator under such scheme would not match
Eq.~\eqref{eq:spectrum-anti}. However, with such discretization scheme, the
continuum limit of determinant $\mathrm{det}[L_f]|_{N \to \infty}$ corresponds
to $\mathrm{det}[\mathcal{L}_f]$ directly. Hence, the ratio of determinant
$\mathrm{det}[L_{f,2\pi}]/ \mathrm{det}[L_{f,0}]|_{N \to \infty} =
\mathrm{det}[\mathcal{L}_{f,2\pi}]/ \mathrm{det}[\mathcal{L}_{f,0}]$.
Moreover, when diagonalizing the transformed operator $H_f=T\cdot L_f$, only
two near zero modes appear around the instanton and anti-instanton in the
phase field trajectory shown in Fig.~\ref{fig:spectrum}(a), a clear signature
of the absence of fermion doubling. Hence, the ratio of determinants following two
different trajectories in Fig.~\ref{fig:spectrum}(a) is predominated by the ratio of these two smallest eigenvalues to the square of the gap, $(E_M/2)^2$.

In summary, we find that associated with a $2\pi$ phase-slip, there is a hidden fermionic zero mode. We can reveal this zero mode by transforming the Lagrangian density operator $L_f$ with $i \sigma^y$ to find a hermitian operator $H_f$. The appearance of the zero mode suppresses $2\pi$ phase-slips. 

\section{Topological Superconducting Devices}
\label{sec:devices}

\begin{figure}
\includegraphics[width=8cm]{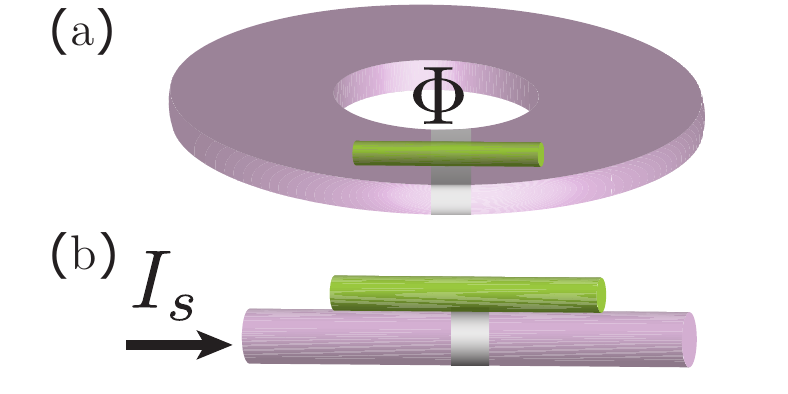} 
\caption{A topological superconducting wire is place across a Josephson junction. In panel (a), the junction is connect to a superconducting ring and a magnetic flux $\Phi$ can be threaded through the ring to bias the conductance energy. In panel (b), the junction is currently biased to form a washboard potential that drives  phase slips.
}
\label{fig:TSC-devices}
\end{figure}

In this section we consider two different setups, which could be built to
detect the suppression of $2\pi$ phase slips experimentally. One setup,
shown in Fig.~\ref{fig:TSC-devices}(a), consists a superconductor ring
interrupted by the Josephson junction while the second setup, shown in
Fig.~\ref{fig:TSC-devices}(b), is a normal Josephson junction with a
constant supercurrent passed through it. While the first setup is conceptually cleaner as the tunneling of a flux quantum out of the loop is measured, the second has the threefold advantage that it does not involve building a loop, that it does not involve an inductance  of a magnitude which is challenging to realize, and that it does not involve changing the inductance $E_L$ but rather the bias current $I_s$ when determining the power-law suppression of the phase-slip rate due to the zero-mode, see below. 

\subsection{Ring geometry} 
\label{sec:ring} 

In the absence of the topological superconductor wire, the Euclidean action of the Josephson junction reads~\cite{makhlin2001}
\begin{multline}
\label{eq:action-phi}
S_{\phi}= \int_{0}^{\beta} d \tau \left[ \frac{1}{2}\frac{1}{8 E_C} \left( \partial_{\tau} \phi(\tau) \right)^2 +E_J( 1- \cos \phi(\tau) ) \right.
\\
\left. + E_L \left(\phi(\tau) - 2\pi \frac{\Phi}{\Phi_0} \right)^2 \right],
\end{multline}
where $\phi$ is the gauge invariant phase difference across the Josephson junction and $\Phi/\Phi_0$ is the ratio of the external magnetic flux threaded through the ring and the superconducting flux quantum $\Phi_0=h/2e$. As a superconductor ring interrupted by a Josephson junction is characterized by its critical current $I_\text{c}$, its capacitance $C$ and the self-inductance $L$ of the ring, we have the following energy scales: the charging energy, $E_C=e^2/ 2 C$, the Josephson energy $E_J= \Phi_0 I_\text{c}/2\pi$ and the inductive energy $E_L = \Phi_0^2/8\pi^2L$.

The potential energy is given by the last two terms of the action Eq.~\eqref{eq:action-phi}. In the absence of the inductance energy as in Eq.~\eqref{eq:sGE}, the cosine potential favors states with $\phi= 2\pi \mathbb{Z}$. The inductance energy breaks such degeneracy by favoring states with $ \phi \approx 2\pi \Phi/\Phi_0$. To still have well defined potential minima at $\phi \approx 2\pi \mathbb{Z}$, we will assume that $E_J$ is the largest energy scale of the system and hence $ E_J \gg E_L$. When $\Phi=0$, there are a global minimum at $\phi=0$ and well defined local minima at $\phi\approx \pm 2 \pi$. As we are interested in the occurrence of phase slips of $2\pi$, i.e., tunneling or relaxation of the phase from one minimum to another, we can first prepare the system with $\Phi=\Phi_0$ at $t<0$ such that a flux quantum is trapped inside the ring and $\phi=2\pi$. Then, we turn off the external flux at $t=0$ and observe the relaxation of phase from $\phi=2\pi$ to $0$ which manisfests itself as voltage spike across the Josephson junction.

As shown in Sec.~\ref{sec:weakLink}, the low energy fermionic degrees of freedom of the topological superconducting wire couple to the gauge invariant phase difference. The effective action is given by
\begin{equation}
\label{eq:action-psi}
S_{\psi}= \int_{0}^{T} d \tau \psi(\tau)^{\dag} \frac{1}{2} \left[ \openone \partial_{\tau} + E_M \cos \phi(\tau)  \sigma^{z} \right] \psi(\tau).
\end{equation}
The presence of fermions influences the tunneling rate between different phase minima. As we showed in Sec.~\ref{sec:weakLink}, the effect of the low energy fermion can be investigated by two routes as detailed below.

\begin{center}
\textbf{1. Integrating out fermions}
\end{center}

Following procedures in Sec.~\ref{subsec:m1}, we can first integrate out the fermionic action Eq.~\eqref{eq:action-psi} and obtain the effective actions for $\Phi=0$
\begin{multline}
\label{eq:action-phi-B+F}
S_\text{eff}^{\pm}= \int_{0}^{\beta} d \tau \left[ \frac{1}{2}\frac{1}{8 E_C} \left( \partial_{\tau} \phi(\tau) \right)^2 +E_J( 1- \cos \phi(\tau) ) \right.
\\
\left. + E_L \phi^2(\tau) \pm \frac{E_M}{2} \cos(\phi(\tau)/2) \right].
\end{multline}
We observe that integrating out of fermionic degrees of freedom simply adds the term $\pm E_M \cos (\phi/2) /2$ into the original bosonic action with the choice of $\pm$ sign depending on the fermion parity of the system. 

\begin{figure}
\includegraphics[width=8.5cm]{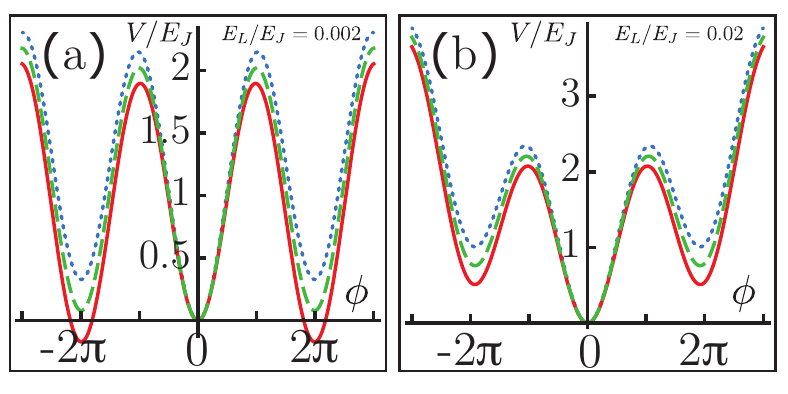}
\caption{The potential profiles of $V^{\pm}(\phi)$ in the action~\eqref{eq:action-phi-B+F} are plotted in solid (red) and dotted (blue) lines for for $V_{\pm}$ respectively, with $E_M/E_J=0.25$. The dashed (green) line is the potential without the Majorana fermions, i.e., $E_M=0$. The panel (a) shows the typical situation for $E_L < E_M/4 \pi^2$ with $E_L/E_J=0.002$, where two degenerate minima sit at $\phi \approx \pm 2\pi$. The panel (b) shows the typical situation for $E_L > E_M/4 \pi^2$ with $E_L/E_J=0.02$, where the potential minimum is at $\phi=0$ and two local minima are around $\phi \approx \pm 2\pi$.}
\label{fig:V-pm}
\end{figure}

To understand the effective actions, we first plot the profiles of the potential term
\be
V^{\pm}(\phi) = E_J( 1- \cos \phi(\tau) ) + E_L \phi^2(\tau) \pm \frac{E_M}{2} \cos(\phi(\tau)/2),
\ee 
in Fig.~\ref{fig:V-pm} with a shift to make all $V^{\pm} (0)=0$. The initial condition is prepared such that the superconducting wire is at its ground state for $\phi=2\pi$. Therefore, with $E_M>0$, the effective action should take the sector $S_\text{eff}^{+}$, which will be assumed throughout the following discussions. We note that the effective potential $V^{+} (\phi)$ behaves qualitatively different depending on $E_L$ is greater or smaller than $E_M/4 \pi^2$. When the inductance energy is dominates, $ E_L > E_M/4 \pi^2$, the potential has a global minimum at $\phi=0$ and two local minima at $\phi=\pm 2\pi$. In contrast, when the Majorana fermion energy becomes substantial, $ E_L < E_M/4 \pi^2$, there are two degenerate minima at $\phi\approx \pm 2\pi$ and a local minimum at $\phi=0$.

From the potential profiles in the $ E_L < E_M/4 \pi^2$ regime, we find that a phase slip from $\phi=2\pi$ to $\phi=0$ is energetically unfavorable as $V^{+} (0) > V^{+}(2\pi)$. Instead, a phase slip of $4 \pi$, tunneling between $\phi=\pm 2\pi$, would lead to a stable state. As discussed earlier, such a phase slip would not change the states of a qubit based on this system.

For the regime where $ E_L > E_M/4 \pi^2$, an initial state at $\phi=2\pi$ can relax to $\phi=0$ state since now $V^{+} (0) < V^{+}(2\pi)$. The relaxation rate is given by $\Gamma_{2\pi\to 0} = K e^{- S'_{0}}$ where $K$ corresponds to the attempt rate for the tunneling and $S'_0$ is the adjusted action evaluated along the bouncing trajectory that starts from the initial energy minimum $\phi_i \approx 2\pi$ to the bouncing point $\phi_b$ and then back to $\phi_i$.~\cite{Coleman1988} Here, the adjusted action is defined by $S'= S_{\text{eff}}^{+} - \int d \tau V^{+}(\phi_i) $ such that the corresponding potential $V'(\phi)= V^{+}(\phi)- V^{+}(\phi_i)$ vanishes at the potential minimum $\phi_i$. As a rough first approximation, we can assume that $K$ is not affected by the presence of Majorana fermions and plays no role for our discussion. 

To compare the relaxation rates, $\Gamma^{M}_{2\pi\to 0}$ (Majorana fermions present) and $\Gamma^{NM}_{2\pi\to 0}$ (Majorana fermions absent), we shall now compute the $S'_0$ for both cases. As the bouncing trajectory is a stationary path of the equation of motion, one can show that
\begin{equation}
\label{eq:S'-0-in-V}
S'_0 
=  \frac{1}{\sqrt{E_C}}  \int_{\phi_i}^{\phi_b} d \phi \sqrt{V'(\phi)}.
\end{equation}
In the case of $E_L = E_M = 0$, we have $\phi_i = 2\pi$ and $\phi_b = 0$, and the action is $S'_{0}=4 \sqrt{2 E_J/E_C}$. When $E_L/E_J\ll 1$, we still have $\phi_i \approx 2\pi$ and $\phi_b \approx 0$, and we can approximate $S'_{0} \approx 4 \sqrt{2 E_J/E_C}$.
Qualitatively, the presence of a small inductance energy $E_L/E_J\ll 1$ increases the relaxation rate only slightly, i.e., decreasing the action such that $S'_{0} \lesssim S'_{0}|_{E_L = 0}$.

We observe that the suppression of tunneling rate due to the Majorana fermions is given by $e^{- \delta S'_0}$, where 
\begin{equation}
\label{eq:delta-S'0}
\delta S'_0 =S'_0 - S'_0|_{E_M = 0},
\end{equation}
is the difference between the actions. From Eq.~\eqref{eq:S'-0-in-V}, one can see that $\delta S'_0$ is of the form $\delta S'_0 =\sqrt{\frac{E_J}{E_C}} f \left(\frac{E_L}{E_J},\frac{E_M}{E_J}\right)$. In the limit $E_J\gg E_L\gg E_M/(4\pi^2)$, one can approximate
\begin{equation}
\label{eq:f-approximation}
f\left(\frac{E_L}{E_J},\frac{E_M}{E_J}\right) \approx \frac{E_M}{2 \sqrt{2} E_J} \ln(E_J/E_L)
\end{equation}
which leads to 
\begin{equation}
\label{eq:delta-S'-approximation}
\delta S'_0 \approx \frac{E_M}{2 \sqrt{2E_C E_J}} \ln(E_J/E_L).
\end{equation}

\begin{figure}
\includegraphics[width=8cm]{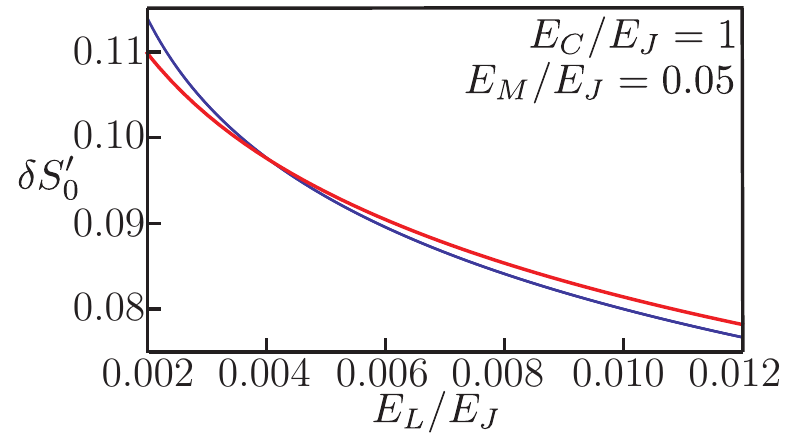}
\caption{The $\delta S'_0$ in Eq.~\eqref{eq:delta-S'0} is evaluated numerically and shown in blue curve as a function of $E_L/E_J$ with $E_C/E_J=1$ and $E_M/E_J=0.05$. The red curve is the approximate result shown in Eq.~\eqref{eq:delta-S'-approximation}.
} 
\label{fig:delta-S'-0}
\end{figure}

It is however straightforward to evaluate $\delta S'_0$ numerically, which is
shown in Fig.~\ref{fig:delta-S'-0} as a function of $E_L/E_J$ with the
parameter $E_M/E_J=0.05$ and $E_C/E_J=1$. The  red line shows the
approximation result in Eq.~\eqref{eq:delta-S'-approximation}. Here, the
positive sign of $\delta S'_0$ indicates the suppression of relaxation rate.
In general, a smaller $E_L/E_J$ and larger $E_M/E_J$ leads to a stronger suppresion. We also note that the approximated form of $f$ only provides a qualitative trend of $f(\frac{E_L}{E_J},\frac{E_M}{E_J})$. However, in the following subsection we will show that the approximate form Eq.~\eqref{eq:f-approximation} is indeed the fingerprint of the zero mode physics.

\begin{center}
\textbf{2. Relation to zero modes}
\end{center}

In the limit that $E_J\gg E_L\gg E_M/(4 \pi^2)$, we can first neglect the presence of the Majorana fermion and follow the bouncing trajectory of action Eq.~\eqref{eq:action-phi}. Then, the Majorana fermion can be integrated out with the assumption that $\phi(\tau)$ follows the bouncing trajectory. Such a trajectory can be evaluated by realizing that
\begin{equation}
\frac{1}{16 E_C} \dot{\phi}^2 - E_J(1-\cos \phi)- E_L \phi^2= E, 
\end{equation} 
is conserved along the classical trajectory. From the initial condition, $\phi = 2\pi$ and $\dot \phi=0$, we have $E= - 4\pi^2 E_L$ and hence the classical trajectory satisfies
\begin{equation}
\label{eq:EOM-phi}
\frac{d \phi}{d t} = 4 \sqrt{E_c (E_J(1-\cos \phi )+E_L \phi^2 -4\pi^2 E_L )}.
\end{equation}

\begin{figure}
\includegraphics[width=8cm]{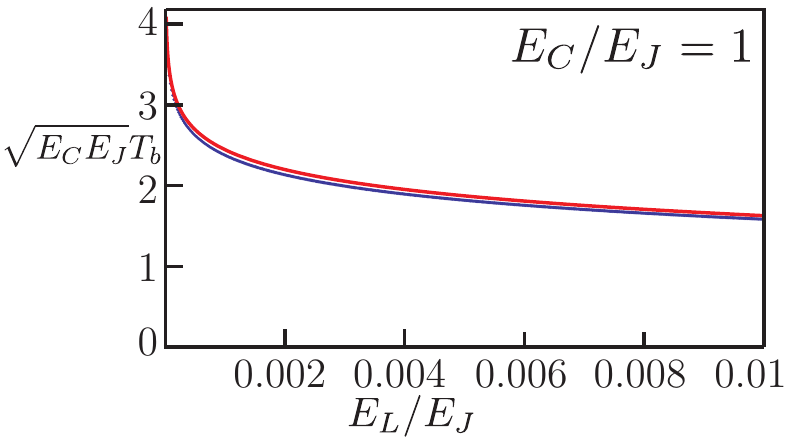}
\caption{The time interval $T_b$ as a function of $E_L/E_J$ is evaluated numerically with $E_C/E_J=1$, c.f. Eq.~\eqref{eq:Tb-integral}, and shown in the blue curve. The red curve is the approximated result in Eq.~\eqref{eq:Tb-approx}.}
\label{fig:Tb-EL}
\end{figure}

As discussed in Sec.~\ref{subsec:m2}, zero modes appear when the
superconductor phase difference $\phi(\tau)$ passes through $\pi$, i.e., from
$\phi>\pi$ to $\phi<\pi$ or vise versa.  For a bouncing event, similar to the
phase trajectory depicted in Fig.~\ref{fig:spectrum}(a), the superconducting phase $\phi(\tau)$ passes through $\pi$ twice, separated by a time interval of $T_b$. Therefore, the zero energy eigenvalues at $\phi=\pi$ split to finite energies $\delta \lambda= \pm E_M e^{-E_M T_b}/2 $. When $E_L> 2 E_J/(3 \pi^2)$ and hence $\phi_b< \pi$, the imaginary time interval of $T_b$ can be readily evaluated from
\begin{equation}
\label{eq:Tb-integral}
 T_b= \frac{2}{4 \sqrt{E_C}} \int_{\phi_b}^{\pi} \frac{d \phi}{ \sqrt{E_J(1-\cos \phi )+E_L \phi^2 -4\pi^2 E_L }}.
\end{equation}
For $E_L/E_J \ll 1$, we can ignore the contributions from $E_L/E_J$ from the integrand. Thus, this integral can be approximated by
\begin{equation}
\label{eq:Tb-integral-approx}
\begin{split}
2 \sqrt{2 E_C E_J} T_b\approx \int_{\phi_b}^{\pi} \frac{d \phi}{|\sin(\phi/2 )| }
= - \ln \tan \frac{\phi_b}{4} 
\end{split}
\end{equation}
with $\phi_{b} \approx 2\pi \sqrt{2 E_L/E_J}$. By droping the constant terms, we have
\begin{equation}
\label{eq:Tb-approx}
 T_b= \frac{1}{ 2 \sqrt{2 E_C E_J} } \ln (E_J/E_L).
\end{equation}
In Fig.~\ref{fig:Tb-EL}, we show the numerically evaluated $T_b$ as a function of $E_L/E_J$ with $E_C=E_J=1$. The approximated $T_b$ in Eq.~\eqref{eq:Tb-approx} is in good agreement with numerical results.

From Fig.~\ref{fig:spectrum}(c), we observe that most eigenvalues remain unchanged in the presence of an instanton despite the appearance of zero modes. As the zero energy modes split to
\begin{equation}
 \delta \lambda= \pm \frac{E_M}{2} (E_L/E_J)^{E_M/(2\sqrt{2E_C E_J })},
\end{equation}
the tunneling rate is changed by the ratio of the determinant of the fermionic kernel in the presence and in the absence of the bounce. This ratio is dominated by
\begin{equation}
\label{eq:decay-rate-effect-fermion}
 \Gamma \propto \frac{\sqrt{\textrm{det}[L_f] }}{\sqrt{\textrm{det}[L_{f,0}] } } \sim \frac{|\delta \lambda|}{E_M/2}= \left(\frac{E_L}{E_J} \right)^{E_M/(2\sqrt{2E_C E_J })}. 
\end{equation}
This result is in perfect agreement of the  suppression of relaxation rate $e^{-\delta S'_0}$ [given in Eq.~\eqref{eq:delta-S'-approximation}] due to the presence of fermions.

Let us now discuss the relevant energy scales and the experimental feasibility of such a system. First, for the bouncing event to cross phase $\pi$, it requires $E_L< 2 E_J/(3 \pi^2)\sim 0.0675 E_J$. We also need $E_L \gg E_M/(4\pi^2)$ to make phase slips of $2\pi$ energetically possible. Therefore, we require a system satisfying the condition $E_J \gg E_L \gg E_M/(4\pi^2)$. Finally, we need $E_M \gtrsim E_C$ to make the dependence on $E_L$ observable as it requires that the exponent in Eq.~\eqref{eq:decay-rate-effect-fermion} is of order unity.

We shall seek an experimental construction with a large $E_J/E_C$ ratio such that the energy scale hierarchy can be realized. In general, a Josephson junction with $E_J \gg E_C$ can be made out of a Nb/AlOx/Nb junction. A typical critical current density of such a junction with insulating layer thickness $1\sim 10$ nm is in the range of $j_c = 10\sim 1000$~A/cm$^2$, see Ref.~\onlinecite{Castellano1996}. For a junction of area $10^{-8}$~cm$^2$ with critical current density $j_c=20$~A/cm$^2$, we can estimate the Josephson energy by $E_J= \Phi_0 I_c/(2\pi)\approx 5$~K. With the thickness of the insulator at $5$ nm, the expected capacitance of such a junction is about $18$ fF and leads to a charging energy at $E_C\approx 200$ mK. For a semiconductor wire in contact with Niobium, $E_M$ can be of the order of $0.1-1$ K as the superconducting critical temperature $T_c\approx 9.2$ K for Niobium. Here, we will assume that $E_M\approx 0.5$ K, which gives the exponent in Eq.~\eqref{eq:decay-rate-effect-fermion} as $E_M/(2\sqrt{2E_C E_J }) \sim 0.18$. Finally, we need a relatively large inductance $L > 12$ nH to satisfy $E_L< 0.0675 E_J$. Such values of inductance can be achieved with a larger ring or with a more complicated design~\cite{Manucharyan2009}. In the following, we will show that the same physics can be accessed in a much simpler setup without inductance at all.

\subsection{Current biased geometry} 
\label{sec:current-bias}

The second geometry we consider is that of a Josephson junction on a topological superconducting wire, and we pass a supercurrent $I_s$ through the wire. The effective action is then
\begin{align}
\label{eq:action-phi-open}
S_\text{eff}&= \int_{0}^{T} d \tau \left[ \frac{1}{2}\frac{1}{8 E_C} \dot{\phi}(\tau)^2 + E_J( 1- \cos \phi(\tau) ) \right. 
\\
&\left. \pm \frac{E_M}{2} \cos(\phi(\tau)/2)+\frac{\hbar}{2e} I_s  \phi \right],
\end{align}
which has a titled doubly periodic washboard potential. 

In the case where there is no supercurrent applied, $I_s=0$, the system relaxes to a stationary state where the superconducting phase difference is pinned to a multiple of $4\pi$. Successively, driving the system with an external current of size $I_s$ tilts the potential. The system is trapped in a metastable state having the possibility to tunnel through the potential barrier out of the local minimum. Employing the same analysis as in the previous subsection, one can show that the presence of the external current plays a similar role as the inductance term in the ring in particular it makes the potential minima separated by $2\pi$ tilted, thus giving the system an incentive to tunnel and thus lower its energy. In particular, the effect of the bias supercurrent on the relaxation rate $\Gamma$ is given by \eqref{eq:decay-rate-effect-fermion} with
\begin{equation}
\label{eq:EL-eff}
E_L \mapsto \frac{\hbar  I_s}{4e \pi } = \frac{\Phi_0 I_s}{4\pi^2}.
\end{equation}
Differently from the previous setup, the phase of the current-bias wire after tunneling enters a so-called running state which means that the wire turns resistive, essentially switching to a normal state~\cite{Fulton1974, Martinis1987}. After turning off the bias current, the superconducting phase retraps in one of the minima due to dissipation given by a small shunt resistor.

The experimental determination of the relaxation rate $\Gamma$ thus
goes along the following line. First, the current-bias is turned off and
the wire is prepared in its ground state. Then, the bias is turned on to a
value $I_s$ on a timescale $T_\text{on} \ll \Gamma^{-1}$. The time difference
between the event of turning on the current bias and the switching of the
voltage to a finite value is a direct measure of the inverse phase-slip
rate $\Gamma^{-1}$.  Repeating the experiments for different values of
$I_s$ the predicted power law \eqref{eq:decay-rate-effect-fermion} can be
tested and thus the suppression of the quantum phase-slip rate due to the
zero-mode when lowering $I_s$ in the regime $4\pi^2 E_J \gg I_s \Phi_0\gg
E_M$ can be confirmed.

Thus far, we have assumed that the initial state (before we turn on the bias supercurrent) corresponds to the Josephson junction localized in the deeper well of the doubly periodic potential. Alternatively, we could prepare the Josephson junction so that it is localized in a random well (e.g. by driving it). With this type of initial condition, there will be two relaxation rates, corresponding to the two types of wells in the doubly periodic potential. Thus, the experimentally observed distribution of waiting times should be bimodal.

\section{Concluding Remarks}
\label{sec:conclusions}
We investigate phase slips in topological superconducting wires. Unlike in conventional superconducting wires, phase slips in topological superconducting wires occur in multiples of $4\pi$ as opposed to multiples of $2\pi$.  Our original motivation for looking into this problem was to understand the effects of phase-slips in topologically protected qubits made up of conventional and topological superconducting wires. As phase-slips are non-local perturbations, they can cause decoherence of a topologically protected qubit. 

The fact that phase-slips in topological wires occur in multiples of $4\pi$ is
well known. Indeed, by integrating out the fermions, one finds that the
effective action for the phase is $4\pi$ periodic. We show an alternative
explanation of this fact by a beautiful analogy to spontaneous symmetry
breaking of the theta vacuum in quantum chromodynamics. For the case of QCD,
t'Hooft found that in the background of the instanton of the gauge field,
there is a zero mode in the fermionic determinant.~\cite{Hooft1976,
Coleman1988} This zero mode results in the vanishing of the transition rate
between configurations of the vacuum with different winding numbers.
Similarly, we find that in the background of a $2\pi$ phase slip, the fermion
determinant contains a ``hidden" zero mode, that results in the vanishing (suppression) of the rate of $2\pi$ phase slips.

Returning to the question of decoherence of qubit devices, we show that phase-slips in the topological wires do not cause decoherence, as they occur in multiples of $4\pi$. However, the qubits are susceptible to decoherence from $2\pi$ phase slips in the conventional superconducting wire segments. Phase-slips near the junction of conventional and topological superconducting segments can also occur in multiples of $2\pi$, and likewise result in decoherence.

To investigate the $4\pi$ periodic nature of topological superconductors we propose two types of experiments. First, we suggest looking for changes of magnetic flux through a ring made up of a topological superconducting wire by multiples of two flux quanta as opposed to one flux quanta for conventional superconducting rings. Second, for a current biased wire, we suggest looking for voltage spikes corresponding to $4\pi$ phase slips as opposed to $2\pi$ phase slips.

\section*{Acknowledgements}

It is our pleasure to thank J. Alicea and G. Refael for useful discussions. The authors
acknowledge support from the Caltech Summer Undergraduate Research Experience
program, the Rose Hills foundation, the Lee A. DuBridge fellowship, the Sherman Fairchild Foundation, 	DARPA-QuEST program, IQIM, the Alexander von Humboldt Foundation, funds of the Erdal \.{I}n\"{o}n\"{u} chair, TUBITAK under grant No. 110T841. IA thanks the Instituut-Lorentz for their hospitality.

\appendix
\section{Weak link model}
\label{app:weakLink}

\begin{figure}
\includegraphics[width=8cm]{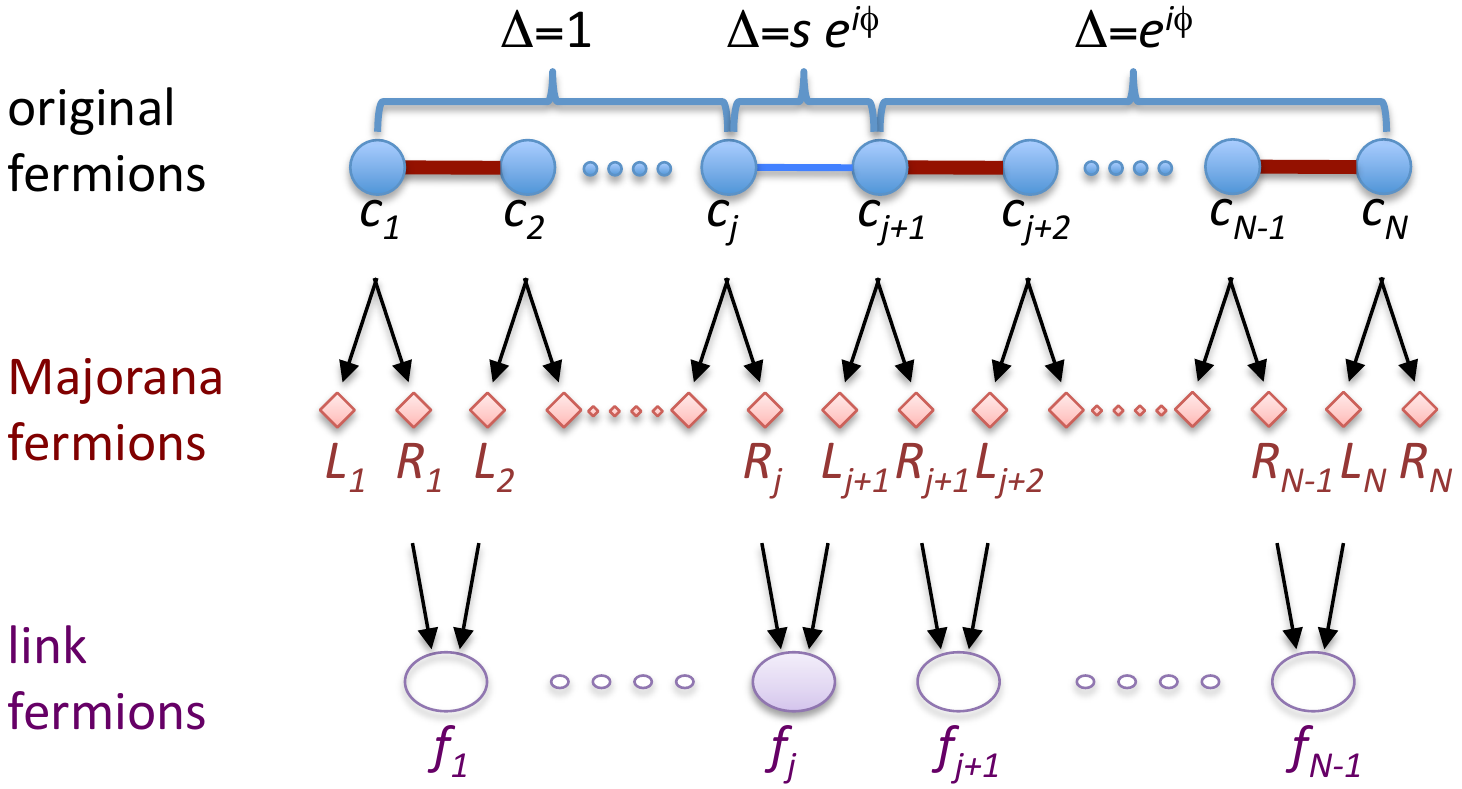}
\caption{Schematic of the Kitaev model extended to contain a weak link. The top row, labeled ``original fermions" shows the original model and indicates the values of the pairing function on the links. The weak link, going from $j$ to $j+1$, is highlighted in blue. The middle row, labeled ``Majorana fermions", indicates the transformation to the Majorana fermion basis. The bottom row, labeled ``link fermions" shows the final transformation to the link fermion basis. The link fermion associated with the weak link is highlighted with purple shading.}
\label{fig:derivationSchematic}
\end{figure}

In this appendix we derive the effective fermionic action for the weak link geometry. We start from the Kitaev model \eqref{H0}, and extend it by letting the magnitude of both $t_{i,i+1}$ and $\Delta_{i,i+1}$ to vary from site to site
\begin{align}
H = & 
- \sum_{i=1}^{N-1}\left( t_{i,i+1} c^{\dagger}_{i+1} c^{\phantom\dagger}_i +  h.c. \right)
\nonumber \\ & \quad\quad\quad
+ \sum_{i=1}^{N-1} \left( \Delta_{i,i+1} c^{\dagger}_{i+1} c^{\dagger}_i +  h.c. \right)
\; .
\end{align}
Next, we set $\Delta_{i,i+1} = t_{i,i+1} e^{i\phi_{i,i+1}}$ on all links. To model the weak link geometry, we need to (1) make a weak link and (2) set the phase of the pairing field to be different to the left and to the right of the weak link. Therefore, we (1) set $t_{i,i+1}=1$ on all links except the weak link, on which we set $t_{j,j+1}=s$ with $s<1$; (2) we set the phase $\phi_{i,i+1}=0$ for $i<j$ and $\phi_{i,i+1}=\phi$ for $i\geq j$, see top row of Fig.~\ref{fig:derivationSchematic}.

Having defined the model, we can obtain the eigenspectrum following the schematic steps illustrated in Fig.~\ref{fig:derivationSchematic}. First, we rewrite the Hamiltonian in slightly more convenient form
\begin{align}
H = &
\sum_{i=1}^{N-1} t_{i,i+1} \left[ 
- c^{\dagger}_{i+1} c^{\phantom\dagger}_i
- c^{\dagger}_i c^{\phantom\dagger}_{i+1}\right.
\nonumber \\ 
&\quad \quad\quad \left.
+ e^{i\phi_{i,i+1}} c^{\dagger}_i c^{\dagger}_{i+1}
+ e^{-i\phi_{i,i+1}} c^{\phantom\dagger}_{i+1} c^{\phantom\dagger}_i\right]\; .
\end{align}
Now we introduce ``right'' ($R = R^{\dagger}$) and ``left'' ($L = L^{\dagger}$) Majorana operators at site $i$ as
\be
c^{\dagger}_i = \frac{e^{-i\phi_{i,i+1}/2}}{2} \left[ R_i + i L_i \right]
\; ,
\ee
equivalent to 
\be
\begin{split} &
R_i = \left[
e^{+i\phi_{i,i+1}/2} c^{\dagger}_i + e^{-i\phi_{i,i+1}/2} c^{\phantom\dagger}_i
\right]
\\ & 
i L_i = \left[
e^{+i\phi_{i,i+1}/2} c^{\dagger}_i - e^{-i\phi_{i,i+1}/2} c^{\phantom\dagger}_i
\right]
\; ,
\end{split}
\ee
which allow us to write
\be\label{H0b}
\begin{split}
H = &
\sum_{i \neq j} 
i R_i L_{i+1} \nonumber\\
&+ i s R_j \left[\sin(\phi/2)R_{j+1}+\cos(\phi/2)L_{j+1}\right]
\; .
\end{split}
\ee
Note at this point that the two Majoranas at the ends of the wire, $L_1$ and $R_N$, do not appear in the 
Hamiltonian - they do not couple to anything, and therefore constitute the two Majorana zero modes at the two ends of the wire. We can recombine the Majorana operators to form fermion operators on the links
\be\label{link_fermions}
f^{\dagger}_i = \frac{1}{2} \left[ R_i - i L_{i+1} \right]
\; ,
\ee
for which 
\be
f^{\dagger}_i f^{\phantom\dagger}_i = \frac{1}{2} \left[ 1 + i R_i L_{i+1} \right]
\; ,
\ee
where we used the fact that $R_i^2 = L_i^2 = \left\{ c^{\phantom\dagger}_i, c^{\dagger}_i \right\} = 1$.
In terms of the link fermions we get
\be
\begin{split}
H = &
\sum_{i\neq j} 
\left( 2 f^{\dagger}_i f^{\phantom\dagger}_i - 1 \right)
\nonumber \\
&+ s \cos(\phi/2)  \left( 2 f^{\dagger}_j f^{\phantom\dagger}_j - 1 \right)
\nonumber \\
&+i s \sin(\phi/2)  \left(f^{\dagger}_j + f^{\phantom\dagger}_j\right)\left(f^{\dagger}_{j+1} + f^{\phantom\dagger}_{j+1}\right)
\; .
\end{split}
\ee
The two Majoranas at the ends of the wire, $L_1$ and $R_N$, may be combined into a single complex fermion $f_0= \frac{1}{2} \left[ R_N - i L_{1} \right]$, which does not appear in the Hamiltonian. 
After transforming to the link operators, we find a Hamiltonian that is almost diagonal. The exception being the terms involving $f_j$ and $f_{j+1}$ operators in the vicinity of the weak link. Explicitly, the non-diagonal part of the Hamiltonian is
\begin{widetext}
\begin{align}
H_{j,j+1}=
\left(
\begin{array}{cccc}
f_j^\dagger & f_j^{\phantom \dagger} & f_{j+1}^\dagger & f_{j+1}^{\phantom \dagger}
\end{array}
\right)
\left(
\begin{array}{cccc}
s \cos(\phi/2) & 0 & i s \sin(\phi/2)/2 & i s \sin(\phi/2)/2\\
0 & - s \cos(\phi/2) & i s \sin(\phi/2)/2 & i s \sin(\phi/2)/2\\
-i s \sin(\phi/2)/2 & - i s \sin(\phi/2)/2 & 1 & 0\\
-i s \sin(\phi/2)/2 & - i s \sin(\phi/2)/2 & 0 & -1
\end{array}
\right)
\left(
\begin{array}{c}
 f_j^{\phantom \dagger} \\
 f_j^\dagger \\
 f_{j+1}^{\phantom \dagger}\\
 f_{j+1}^\dagger 
\end{array}
\right).
\label{eq:Hnd}
\end{align}
\end{widetext}
The non-diagonal part of the Hamiltonian may be readily diagonalized via a Bogoliubov transformation with eigenvalues
\begin{align}
\epsilon=\pm\sqrt{\frac{1}{2}\left(1+s^2\mp \sqrt{1+s^4-2s^2 \cos(\phi)}\right)}.
\end{align}
Here, we denote two positive eigenvalues (at $\phi=0$) to correspond to the annihilation operators $c_{w}^{\phantom \dagger}$ and $d_{w^{\phantom \dagger}}$ and the two negative eigenvalues to the creation operators $c_{w}^{\dagger}$ and $d_{w}^{\dagger}$. Away from $\phi=0$, we label the operators such that their eigenvalues evolve smoothly. Putting all these considerations together, the Hamiltonian becomes
\begin{align}
H&=\sum_{i=1..j-1,j+2..N-1} \left( 2 f^{\dagger}_i f^{\phantom\dagger}_i - 1 \right) \label{eq:Hfull}\\
&+ 0 \times \left( 2 f^{\dagger}_0 f^{\phantom\dagger}_0 - 1 \right) \nonumber\\
&+\sqrt{\frac{1+s^2 - \sqrt{1+s^4-2s^2 \cos(\phi)}}{2}} 
\left( 2 c^{\dagger}_w c^{\phantom\dagger}_w - 1 \right) \nonumber\\
&+\sqrt{\frac{1+s^2 + \sqrt{1+s^4-2s^2 \cos(\phi)}}{2}}
\left( 2 d^{\dagger}_{w} d^{\phantom\dagger}_{w} - 1 \right)\,, \nonumber
\end{align}
where the first line corresponds to all the link fermions except $f_j$ and $f_{j+1}$, the second line corresponds to the zero mode on the ends of the wire, and the final two lines represent the link fermions, $f_j$ and $f_{j+1}$, around the weak link in their diagonalized basis.

\begin{figure*}
\centering
\subfigure[$\,s=0.25$]{
\includegraphics[scale=0.55]{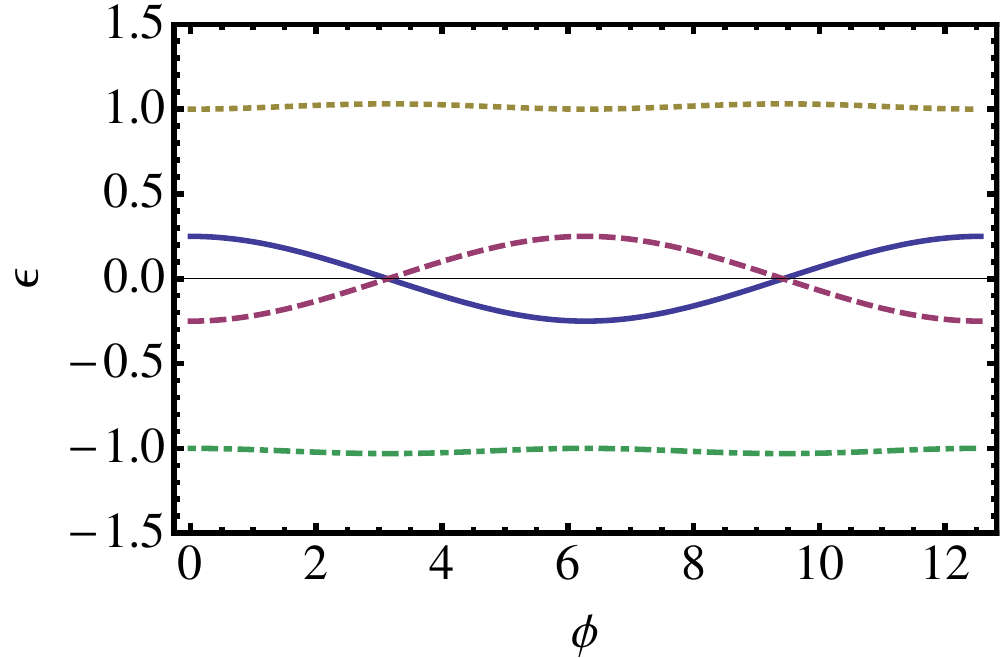}
\label{fig:subfig1}
}
\subfigure[$\,s=0.75$]{
\includegraphics[scale=0.55]{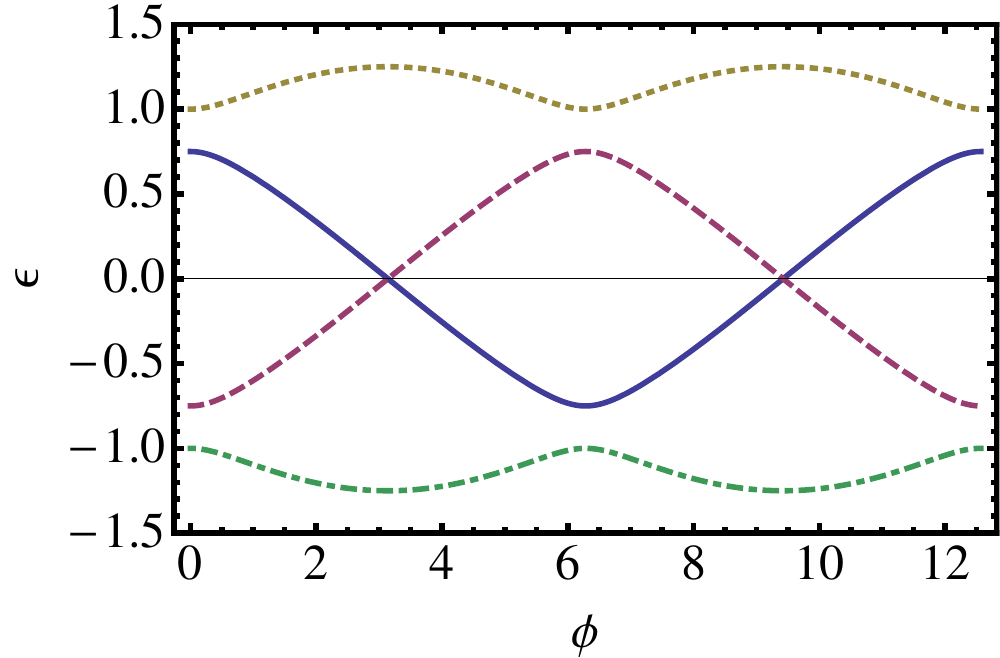}
\label{fig:subfig2}
}
\subfigure[$\,s=1.0$]{
\includegraphics[scale=0.55]{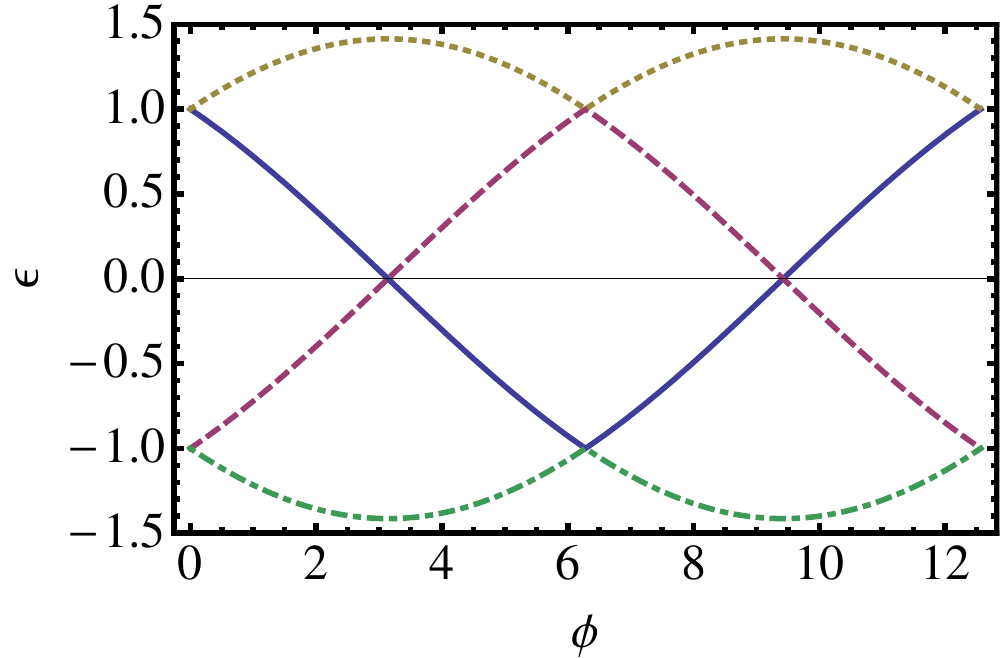}
\label{fig:subfig3}
}
\caption{Spectrum of the four eigenvalues of \eqref{eq:Hnd} as a function of the phase difference $\phi$ across the weak link for three values of $s$ parameter as indicated. The traces correspond to adiabatic evolution of the eigenvalues, two eigenvalues indeed cross at $\pi$ and $3\pi$. Thus we see that the spectrum has both a $2\pi$ periodic component (the two eigenvalues at $\epsilon \sim \pm 1$) as well as a $4\pi$ periodic component (the two eigenvalues at $\epsilon \sim 0$).}
\label{fig:Kitaev_wire_spectrum}
\end{figure*}

To better understand the fermions on the links $j,j+1$ and $j+1,j+2$, we plot the eigenvalues as a function of the phase difference across the weak link in Fig.~\ref{fig:Kitaev_wire_spectrum}. The figure shows two distinct types of eigenvalues: (1) The eigenvalues correspond to $c_w$ are near zero energy with eigenfunction localized mostly on the link $j,j+1$. (2) The eigenvalues correspond to $d_w$ are near $\pm1$ with eigenfunction localized mostly on the link $j+1,j+2$. Case (1) corresponds to the eigenvalues that cross zero at $\pi$ and $3\pi$. The crossing indicates that the creation and annihilation operators $c_w^{\phantom \dagger}$ and $c_w^\dagger$ switch after the phase rotates by $2\pi$. As we make the weak link stronger (by increasing $s$), the two types of eigenvalues approach each other at $\phi=0,2\pi,4\pi$. In the limit $s=1$, we return to the case of a continuous wire, and the eigenvalues touch at multiples of $2\pi$.

Using our understanding of the adiabatic evolution of the eigenvalues, we can choose the correct branch of the square root in writing the penultimate term of the Hamiltonian \eqref{eq:Hfull} (i.e. the solid line in Fig.~\ref{fig:Kitaev_wire_spectrum}). Thus, expanding to second order in $s$, we can write the Hamiltonian as
\begin{align}
H&=\sum_{i=1..j-1,j+2..N-1} \left( 2 f^{\dagger}_i f^{\phantom\dagger}_i - 1 \right) \label{eq:Hfull2}\\
&+ 0 \times \left( 2 f^{\dagger}_0 f^{\phantom\dagger}_0 - 1 \right) \nonumber\\
&+s \cos(\phi/2)
\left( 2 c_w^{\dagger} c_w^{\phantom\dagger} - 1 \right) \nonumber\\
&+\left(1+ \frac{s^2}{4}\left[1-\cos(\phi)\right]\right)
\left( 2 d^{\dagger}_{w} d^{\phantom\dagger}_{w} - 1 \right)\,. \nonumber
\end{align}
What does the ground state of this Hamiltonian look like? Starting from the state $|0\rangle$ with no $c_i$ fermions we obtain two degenerate ground states
\begin{align}
|GS_1\rangle & = f_0 c_w d_{w} \prod_i f_i |0\rangle \\
|GS_2\rangle & = c_w d_{w} \prod_i f_i |0\rangle
\end{align}
corresponding to the zero mode at the ends of the wire filled and empty. 

Thus far, we have implicitly assumed that our operators are phase dependent, i.e. $c_w=c_w(\phi)$. To understand the evolution of the ground state, we can relate the phase evolved operators to the operators at $\phi=0$. Under the rotation of the phase $\phi$: $0 \rightarrow 2\pi$ the operators (as defined at $\phi=0$) evolve as: 
\begin{align}
f_i(0)  &\rightarrow \left\{
\begin{array}{cc}
f_i(0) & i<j\\
-f_i(0) & i>j+1
\end{array}\right.\\
f_0(0) &\rightarrow -f_0^\dagger(0)=f_0(2\pi)\\
c_w(0) &\rightarrow c_w^\dagger(0)=c_w(2\pi)\\
d_w(0) &\rightarrow -d_{w}(0)
\end{align}
Using these rules we can easily work out the evolution of the ground state:
\begin{align}
|GS_1\rangle&\rightarrow (-1)^{N-j} f_0^\dagger(0) c_w^\dagger(0) d_{w}(0) \prod_i f_i(0) |0\rangle\; , \\
|GS_2\rangle&\rightarrow (-1)^{N-j-1} c_w^\dagger(0) d_{w}(0) \prod_i f_i(0) |0\rangle\; .
\end{align}
Applying the operator transformation rules twice, we find that a $4\pi$ phase slip returns the system to the original configuration.

\begin{figure}
\includegraphics[width=8cm]{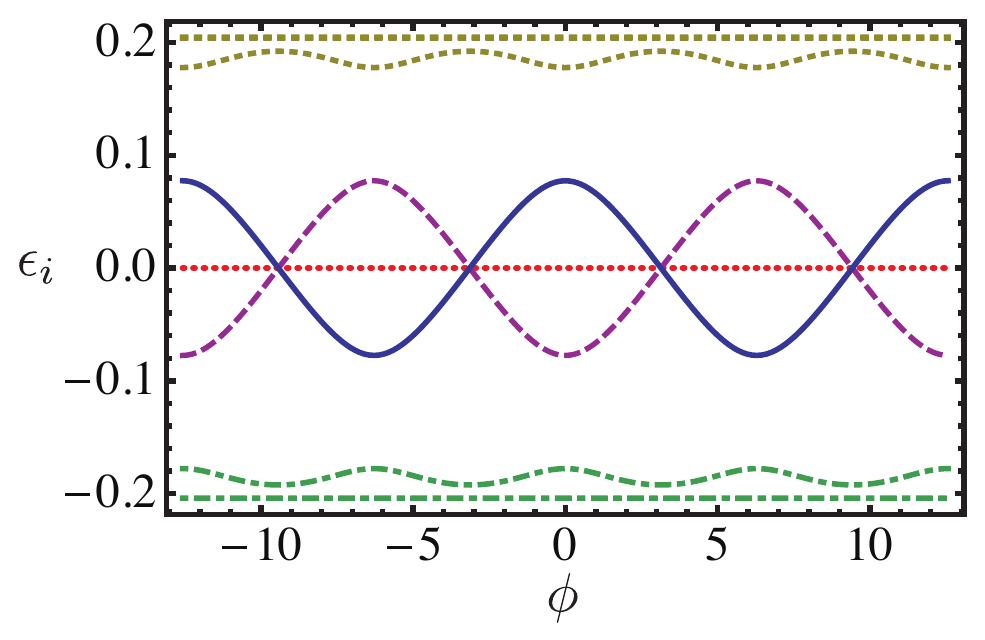}
\caption{Generic fermion spectrum as a function of the phase slip angle in a topological wire showing three types of modes: (1) zero mode associated with the Majorana fermions at the ends of the wire (red dotted line along the horizontal axis), (2) $4\pi$ periodic mode associated with the fermion localized on the phase slip, and (3) a set of $2\pi$ periodic modes associated with the un-localized fermions.}
\label{fig:genericSpectrum}
\end{figure}

Thus far, we have obtained the effective Hamiltonian for the very special case of the Kitaev model at $|\Delta|=t$. However, we point out that the form of the Hamiltonian, and of its spectrum, is generic: containing three ingredients (1) a zero mode associated with Majoranas at the ends of the wire (2) a single mode that corresponds to a Fermion localized on the weak link that is $4\pi$ periodic, and (3) a set of $2\pi$ periodic modes. In particular, this form of the spectrum is not sensitive to adding small disorder, the relaxation of the condition $|\Delta|=t$, nor the spreading out of the ``weak link" over several links of the model. We plot the spectrum for several generic cases in Fig.~\ref{fig:genericSpectrum}.

To understand the evolution, we need to obtain the effective action. The imaginary time (Euclidean) action for the Kitaev model is
\be
S = \int_0^{\beta} d\tau \left[ \sum_{i=1}^{N} c^{\dagger}_i(\tau) \partial_{\tau} c^{\phantom\dagger}_i(\tau) - H \right]
\; .
\ee
We now focus on the first term, and perform on it the same transformation that we used to diagonalize the Hamiltonian. Transforming first to the Majorana fields, we have
\be
\begin{split}
c^{\dagger}_i(\tau) \partial_{\tau} c^{\phantom\dagger}_i(\tau) & = 
\frac{e^{-i\phi/2}}{2} \left[ R_i + i L_i \right]
\partial_{\tau}
\frac{e^{+i\phi/2}}{2} \left[ R_i - i L_i \right]
\\ & = 
\frac{1}{4} \Big[ 
R_i \partial_{\tau} R_i
+
L_i \partial_{\tau} L_i
+
i L_i \partial_{\tau} R_i
-
i R_i \partial_{\tau} L_i
\\ & 
+
i\partial_{\tau} \phi
(1 + i L_i R_i)
\Big]
\; ,
\end{split}
\ee
where we have allowed for the possibility that the superconducting phase $\phi$ changes with (imaginary) time.
Using the anticommutation of Majorana fields, and integration by parts we find
\be\label{iden1}
L \partial_{\tau} R - R \partial_{\tau} L =
L \partial_{\tau} R + (\partial_{\tau} R) L =
\left\{ L , \partial_{\tau} R \right\} = 0
\; ,
\ee
from which we get to
\be
\begin{split}
c^{\dagger}_i(\tau) \partial_{\tau} c^{\phantom\dagger}_i(\tau) & = 
\frac{1}{4} \Big[ 
R_i \partial_{\tau} R_i
+
L_i \partial_{\tau} L_i
+
i\partial_{\tau} \phi
(1 + i L_i R_i)
\Big]
\; .
\end{split}
\ee
Next we want to change fields to the link fermions. Using \eqref{iden1} again, we can show that
\be
f^{\dagger}_i(\tau) \partial_{\tau} f^{\phantom\dagger}_i(\tau) = 
\frac{1}{4} \left[ 
R_i \partial_{\tau} R_i
+
L_{i+1} \partial_{\tau} L_{i+1}
\right]
\; ,
\ee
which yields
\be\label{dynamical_term}
c^{\dagger}_i(\tau) \partial_{\tau} c^{\phantom\dagger}_i(\tau) = 
f^{\dagger}_i(\tau) \partial_{\tau} f^{\phantom\dagger}_i(\tau) + 
\frac{1}{4} i\partial_{\tau} \phi
(1 + i L_i R_i)
\; .
\ee
Now we also need
\be
i L_i R_i = \left[ f^{\phantom\dagger}_{i+1} - f^{\dagger}_{i+1} \right] \left[ f^{\phantom\dagger}_i + f^{\dagger}_i \right]
\; ,
\ee
where we have used
\be
\begin{split} &
R_i = f^{\phantom\dagger}_i + f^{\dagger}_i
\\ & 
i L_{i+1} = f^{\phantom\dagger}_i - f^{\dagger}_i
\; .
\end{split}
\ee
We see that allowing the phase $\phi$ to change dynamically introduces an off-diagonal term 
$\frac{1}{4} i \partial_\tau \phi \left[ f^{\phantom\dagger}_{i+1} - f^{\dagger}_{i+1} \right] \left[ f^{\phantom\dagger}_i + f^{\dagger}_i \right]$
in the link fermion action that corresponds to exciting pairs of link fermions on the neighboring sites. If $\partial_{\tau} \phi$ is much smaller than the system gap, this process will be forbidden, and this term can be safely ignored. We shall assume that we are indeed working in this limit (e.g. due to extrinsic factors like the capacitance that dictates how fast the phase can evolve).  Although the off-diagonal term has no effect on link fermions away from the weak link, what about the fermions $c_w$ and $d_{w}$? Applying the transformation used to diagonalize Eq.~\eqref{eq:Hnd}, we find that we must again excite pairs of quasi-particles (i.e. terms of the form $c_{w}^\dagger d_{w}^\dagger$), which is again forbidden by energy conservation. 

Combining these considerations, and integrating over all link fermions except the $c_w$ fermion, we arrive at the effective action for the weak link
\begin{align}
S_\text{eff}= c_w^\dagger \partial_\tau c_w - s \cos(\phi/2) (2 c_w^\dagger c_w-1)
\\
+ \left(1+ \frac{s^2}{4} \left[1-\cos(\phi)\right]\right)\; . \nonumber
\end{align}

\section{Discretization of fermionic kernel}
\label{app:disc}

In this Appendix, we establish an alternative discretization scheme that avoids the second fermion doubling for the Lagrangian density
\begin{equation}
\label{eq:fermion-kernel}
 \mathcal{L} _f=
\left(
\begin{array}{cc}
\partial_\tau + \Delta (\tau) & 0 \\ 
0 & \partial_\tau - \Delta (\tau) 
\end{array}  
\right),
\end{equation}
in Eq.~\eqref{eq:Lf}, where $\Delta (\tau) = E_M \cos \phi(\tau)$. Hence, the differential equation,
\begin{equation}
\label{eq:kernel-EOM}
\left( \openone \partial_\tau + \sigma_z \Delta(\tau)  \right)
\left(
\begin{array}{c}
u(\tau)
\\
v(\tau)
\end{array}
\right)
=
\lambda
\left(
\begin{array}{c}
u(\tau)
\\
v(\tau)
\end{array}
\right)
\end{equation}
can be solved numerically. Then, the determinant of the kernel is simply given by the product of all eigenvalues, $\lambda$, in the range of $\tau=(0,\beta)$ with $\beta \to \infty$.

The fermionic kernel appearing in the eigenvalue equation, Eq.~\eqref{eq:kernel-EOM}, can be brought into a hermitian form by multiplying from the left with $-i \sigma_y$, namely $H_f= -i \sigma_y \mathcal{L}_f$. This results in the eigenvalue problem
\begin{equation}
\label{eq:poly-EOM}
\left( -i \sigma_y \partial_\tau + \sigma_x \Delta(\tau)  \right)
\left(
\begin{array}{c}
u'(\tau)
\\
v'(\tau)
\end{array}
\right)
=
\lambda'
\left(
\begin{array}{c}
u'(\tau)
\\
v'(\tau)
\end{array}
\right). 
\end{equation}
Such differential equation, in turn, describes the continuous theory of the polyacetylene with identification of $\tau$ as the one-dimensional coordinate and $\Delta(\tau)$ as the mass field.~\cite{Su1979} As $\mathrm{det}[i \sigma_y]=1$, we have $\mathrm{det}[\mathcal{L}_f]= \mathrm{det}[H_f]= \prod \lambda'$.

It is well known that a zero energy mode in $H_f$ will appear and localize at a soliton where the mass field changes sign. In our system, the mass field changes its sign when the superconductor phase alters from $\phi > \pi$ to $\phi < \pi$ or vice versa. Therefore, we expect a vanishing determinant, as one of $\lambda'$ is zero, when such phase change occurs in the imaginary time range $\tau = (0, \beta)$. The appearance of such zero energy modes, however, becomes illusive in Eq.~\eqref{eq:kernel-EOM}. Hence, a proper discretization scheme to make it a difference equation becomes handy for understanding the effect of those zero modes.

Let us first review how to obtain the continuum limit of Polyacetylene, Eq.~\eqref{eq:poly-EOM}, from the lattice model. Then, the discretization scheme for Eq.~\eqref{eq:kernel-EOM} can be inferred by reverse engineering the corresponding parameters. Here, we will follow closely the derivation in Ref.~\onlinecite{Jackiw1983} and assume open boundary conditions.

We consider the following one-dimensional lattice Hamiltonian on $2N$ sites
\begin{multline}
\label{eq:H-poly-1}
H_p= - \sum_{j=2n-1} t_{j+1,j} \left(  {a_{j}}^{\dag} b_{j+1} +{b_{j+1}}^{\dag} a_{j}  \right)
\\
- \sum_{j=2n} t_{j+1,j} \left(  {b_{j}}^{\dag} a_{j+1} +{a_{j+1}}^{\dag} b_{j}  \right),
\end{multline}
where $a_j({a_j}^{\dag})$ and $b_j({b_j}^{\dag})$ are annihilation(creation) operators at the odd and even lattice sites, respectively, and $n\geq 1$ are integers. Here, $t_{j+1,j}$ are real hopping amplitudes between site $j$ and $j+1$ and take the form
\begin{equation}
t_{j+1,j}= t_0 - \gamma (y_{j+1}-y_{j}),
\end{equation}
where $y_j$ is the displacement of the atoms at site $j$ and $\gamma$ is a constant governing the variation of the hopping strength due to the displacement. The lattice constant is $\delta/2$ and hence the distance between two adjacent $a$ (or $b$) type atoms is $\delta$. As the lattice displacements reflects the induced Peierls instability, they take the form $y_j=(-1)^{j} \eta_{j}$. Then, the hopping amplitudes become
\begin{equation}
\label{eq:hop-amp}
t_{j+1,j}= t_0 + (-1)^{j} \gamma (\eta_{j+1} + \eta_{j}).
\end{equation}
Here, $\eta_j$ can be a function of space, but if $\eta_j$ are uniform the hopping amplitudes are simply alternating between even and odd sites. 

To match the continuum theory with Eq.~\eqref{eq:poly-EOM}, it is convenient to introduce the gauge transformation
\begin{equation}
\label{eq:gauge-trans}
a_{j} \to (-1)^{\frac{j+1}{2}} (i)e^{-i\pi/4} a_{j}, \quad b_{j} \to (-1)^{j/2}e^{i \pi/4} b_{j}.
\end{equation}
and then the lattice Hamiltonian becomes
\begin{multline}
\label{eq:H-poly-2}
H_p= -\sum_{j=2n-1} t_{j+1,j} \left(  {a_{j}}^{\dag} b_{j+1} +{b_{j+1}}^{\dag} a_{j}  \right)
\\
+ \sum_{j=2n} t_{j+1,j} \left(  {b_{j}}^{\dag} a_{j+1} +{a_{j+1}}^{\dag} b_{j}  \right).
\end{multline}
We note that Eq.~\eqref{eq:H-poly-1} and Eq.~\eqref{eq:H-poly-2} differ only by the gauge transformation and have exactly the same spectrum.

By arranging operators with $2N$ lattice sites in a vector form
\begin{equation}
\Psi=\left(a_{1},a_{3}, \dots, a_{2N-1} , b_{2}, b_{4},\dots,b_{2N} \right)^{T},
\end{equation}
the lattice Hamiltonian becomes
\begin{equation}
\label{eq:H-poly-3}
H_p= \Psi^{\dag} \mathcal{H}_L \Psi, \quad \mathcal{H}_L = 
\left(
\begin{array}{cc}
0 & Q
\\
Q^{\dag} & 0
\end{array}
\right),
\end{equation}
where $Q$ is an $N \times N$ matrix and has the form
\begin{multline}
Q=
\left(
\begin{array}{ccccc}
- t_0 & 0 & 0 & 0 &\cdots
\\
t_0 & -t_0 & 0 &0 & \cdots
\\
0 & t_0 & -t_0 & 0 &\cdots
\\
0 & 0 & t_0 & -t_0 & \cdots
\\
\vdots & \vdots &  \vdots &\ddots & \ddots
\end{array}
\right)
\\
+
\gamma
\left(
\begin{array}{ccccc}
 \eta_{2}+\eta_{1} & 0 & 0 & 0 &\cdots
\\
\eta_{3}+\eta_{2} & \eta_{4}+\eta_{3} & 0 &0 & \cdots
\\
0 & \eta_{5}+\eta_{4} & \eta_{6}+\eta_{5} & 0 &\cdots
\\
0 & 0 & \eta_{7}+\eta_{6} & \eta_{8}+\eta_{7} & \cdots
\\
\vdots & \vdots &  \vdots &\ddots & \ddots
\end{array}
\right).
\end{multline}
In what follows, we will show that the first sector of $Q$ corresponds to kinetic energy while the second sector of $Q$ corresponds to the mass field.

\begin{figure}
\includegraphics[width=8cm]{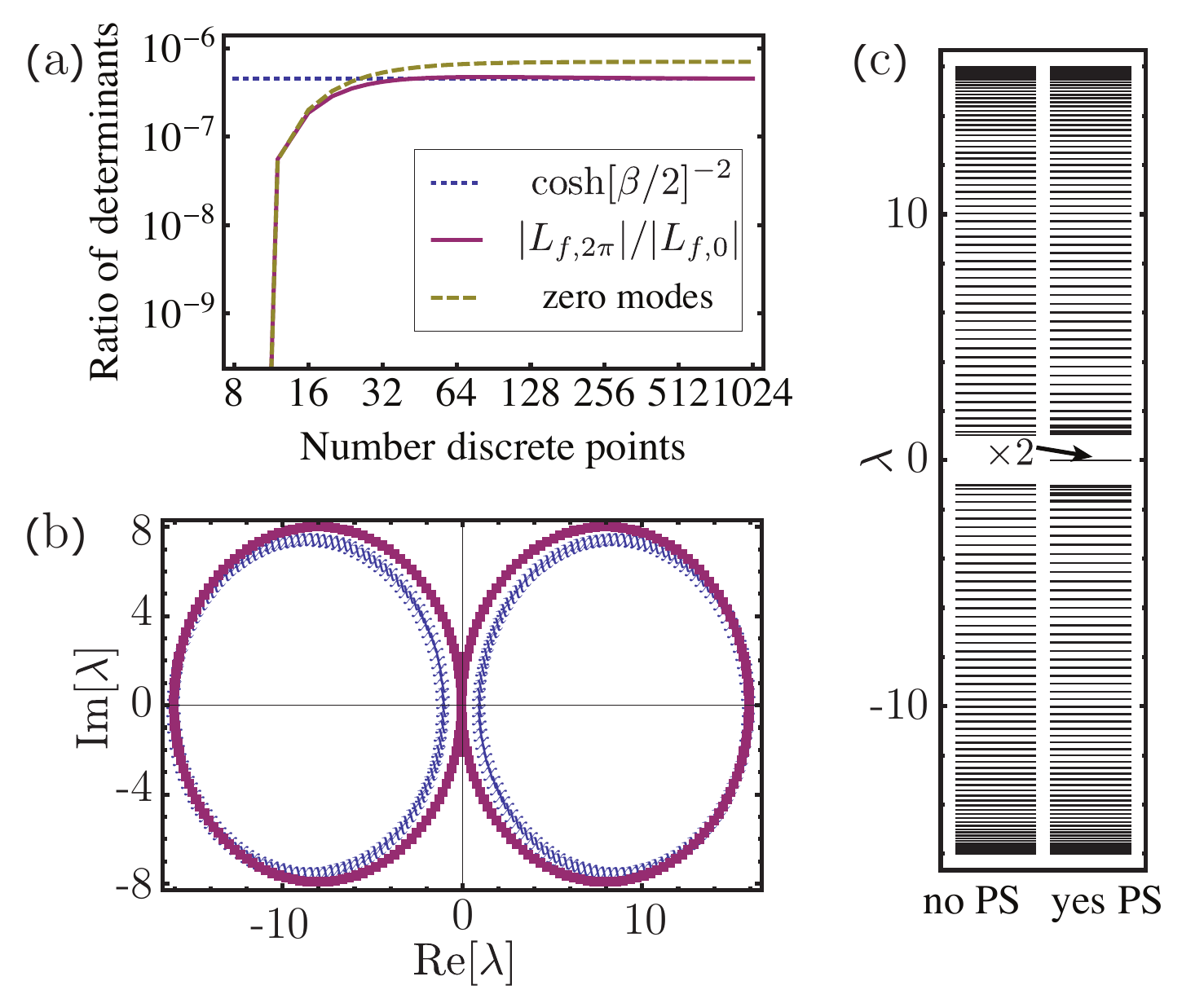}
\caption{
Spectrum using the modified discretization scheme Eq.~\eqref{eq:lattice-iden-kernel}, that avoids the additional fermion doubling.
(a) Ratio of determinant, same as Fig.~\ref{fig:det}, using modified discretization scheme.
(b) and (c) same as (b) and (c) of Fig.~\ref{fig:spectrum}, using modified discretization scheme.
}
\label{fig:spectrumMOD}
\end{figure}

To derive the continuum theory of Hamiltonian \eqref{eq:H-poly-2}, we first identify the operators with continuums fields 
\begin{equation}
 a_{j} = (\delta)^{1/2} U\left((j+1)\frac{\delta}{2} \right), \; b_{j} = (\delta)^{1/2} V\left(j\frac{\delta}{2}\right).
\end{equation}
Inserting this into Eq.~\eqref{eq:H-poly-2}, we have
\begin{equation}
\nonumber
\begin{split}
H_p= & + t_0 \delta \sum_{n} V^{\dag} [n \delta] (U[(n+1)\delta ] -U[n \delta]) 
\\
&- t_0 \delta \sum_{n} U^{\dag} [n \delta] (V[ n\delta ] -V[(n-1) \delta]) 
\\
& + \gamma \delta\sum_{n} (\eta_{2n} + \eta_{2n-1}) V^{\dag}[n \delta] U[n \delta] + h.c.
\\
&+ \gamma \delta\sum_{n} (\eta_{2n+1} + \eta_{2n}) V^{\dag}[n \delta] U[(n+1) \delta]+ h.c.
\end{split}
\end{equation}
If we assume that all $U$, $V$, and $\eta$ are smooth functions in $x$-direction and expand the Hamiltonian to the linear order in $\delta$, the Hamiltonian becomes
\begin{equation}
\begin{split}
H_p= & + t_0 \delta^2 \sum_{n} V^{\dag} (n \delta)  U'(n\delta ) -  U^{\dag} (n \delta) V'(n \delta)
\\
& + 4 \gamma \delta\sum_{n} \eta(n \delta) V^{\dag}(n \delta) U(n \delta) + h.c.,
\end{split}
\end{equation}
where
\begin{equation}
\begin{split}
U'(n\delta) \equiv &(U[(n+1)\delta ] -U[n \delta])/\delta
\\
V'(n\delta) \equiv & (V[ n\delta ] -V[(n-1) \delta])/\delta .
\end{split}
\end{equation}

By taking the continuum limit, we replace $\delta \sum_{n} \to \int dx$, $U(n \delta)\to U(x)$, $V(n \delta) \to V(x)$, $V'(n\delta)\to \partial_x U(x)$,  $V'(n\delta) \to \partial_x V(x)$ and $\eta(n\delta) \to \eta(x)$, and obtain 
\begin{equation}
\begin{split}
H_p= & + t_0 \delta \int dx V^{\dag} (x) \partial U(x) -  U^{\dag} (x) \partial V(x)
\\
& + 4 \gamma \int dx \eta(x) (V^{\dag}(x) U(x) + U^{\dag}(x) V(x) ).
\end{split}
\end{equation}
The continuum Hamiltonian can be rewritten with the vector operators $\chi(x)= (U(x), V(x))^{T}$ as
\begin{equation}
H_p=  \int dx  \chi^{\dag} (x) [ (t_0 \delta)  (-i \partial)\sigma_y  + 4 \gamma \eta(x) \sigma_x ] \chi(x).
\end{equation}
This matches the form of Eq.~\eqref{eq:poly-EOM} with
\begin{equation}
\label{eq:matching-con}
\begin{split}
t_{0} \delta = 1  \quad \to & \quad  t_0 = 1/\delta,
\\
4\gamma \eta(x)= \Delta(x) \quad \to & \quad \gamma \eta(x)= \Delta(x)/4 .
\end{split}
\end{equation}

By defining $\gamma \eta_{j} = \Delta (j \frac{\delta}{2})/4\equiv \Delta_j/4$, we have the lattice identification of the continuum operator $\mathcal{H}=-i \partial_x \sigma_y  + \Delta(x) \sigma_x $ as $\mathcal{H}_L$ in Eq.~\eqref{eq:H-poly-3} and
\begin{multline}
Q= \frac{1}{\delta}
\left(
\begin{array}{cccc}
- 1 & 0 & 0  &\cdots
\\
1 & -1 & 0 &  \cdots
\\
0 & 1 & -1  & \cdots
\\
\vdots & \vdots &\ddots & \ddots
\end{array}
\right)
\\
+
\frac{1}{4}
\left(
\begin{array}{cccc}
 \Delta_{2}+\Delta_{1} & 0 & 0  &\cdots
\\
 \Delta_{3}+ \Delta_{2} & \Delta_{4}+ \Delta_{3} & 0 & \cdots
\\
0 &  \Delta_{5}+ \Delta_{4} &  \Delta_{6}+ \Delta_{5} &\cdots
\\
\vdots & \vdots  &\ddots & \ddots
\end{array}
\right).
\end{multline}
We notice that the discretization of the mass field is spread out over several matrix elements. As Eq.~\eqref{eq:kernel-EOM} and Eq.~\eqref{eq:poly-EOM} differ by a multiplication of $-i \sigma_{y}$, the difference operator of kernel $\mathcal{L}$ is given by
\begin{equation}
\label{eq:lattice-iden-kernel}
L_f = (i \sigma_y \otimes \openone_N) \mathcal{H}_L = 
\left(
\begin{array}{cc}
Q^{\dag} &0
\\
0 & -Q 
\end{array}
\right)
\end{equation}
where $\openone_N$ is a $N\times N$ identity matrix. 

We compare the results for fermions in the background of a phase-slip followed
by an anti-phase slip using the original discretization scheme of
Sec.~\ref{sec:weakLink} and the alternative discretization scheme discussed in
this appendix. Using the alternative discretization scheme, we find that in
the continuum limit the value of the fermion determinant after taking a square
root (Fig.~\ref{fig:spectrumMOD}(a)) is identical to the one obtained in the
original scheme after taking the fourth root (Fig.~\ref{fig:det}). However,
the spectrum of the untransformed Lagrangian (Fig.~\ref{fig:spectrumMOD}(b))
does not have the nice properties found in the original scheme
(Fig.~\ref{fig:spectrum}(b)) due to the explicit mixing of the left and right
moving components. The number of hidden zero modes indeed decreases from four
(Fig.~\ref{fig:spectrum}(c)) in the original scheme to two
(Fig.~\ref{fig:spectrumMOD}(c)) in the alternative discretization scheme.

\bibliography{phaseSlip}
\bibliographystyle{apsrev}

\end{document}